\documentclass[12pt,a4paper]{article}

\usepackage{authblk}

\usepackage{booktabs}
\usepackage{mathptmx}

\usepackage{amsmath,amsfonts,latexsym,amssymb}
\usepackage{verbatim,amsthm,curves,graphics}
\usepackage{mathrsfs}
\usepackage[dvips]{graphicx}
 
\usepackage{todonotes,soul} 
 
\numberwithin{equation}{section}
 
\usepackage{latexsym}
\usepackage{mathrsfs} 
\usepackage{epsfig}
\usepackage{longtable}
\usepackage{slashed}
\usepackage[all]{xy}
\usepackage{graphicx,tikz}
\usetikzlibrary{arrows}
\usepackage{setspace}

\usepackage{graphicx,color}
\usepackage{array}
\usepackage{tikz}
\usetikzlibrary{shapes.geometric}
\usetikzlibrary{decorations.pathmorphing}
\usetikzlibrary{arrows,shapes,positioning}
\usetikzlibrary{decorations.markings}
\tikzstyle arrowstyle=[scale=1]
\tikzstyle directed=[postaction={decorate,decoration={markings,mark=at position .65 with {\arrow[arrowstyle]{stealth}}}}]
\tikzstyle reverse directed=[postaction={decorate,decoration={markings,mark=at position .65 with {\arrowreversed[arrowstyle]{stealth};}}}]
\def\ben{\begin{equation}}
\def\een{\end{equation}}
\def\bena{\begin{eqnarray}}
\def\eena{\end{eqnarray}}
\def\non{\nonumber}

\def\d{{\rm d}}

\def\L{{\cal L}}

\def\t{\frac{t}{2}}

\def\path{\operatorname{Pa}}

\def\H{\mathcal{H}}
\newcommand{\ctn}{{\rm ctg}}

\newcommand{\myid}{{\bf 1}}
\newcommand{\HH}{{\mathbb H}}
\renewcommand{\SS}{{\mathbb S}}
\newcommand{\DD}{{\mathbb D}}
\newcommand{\RR}{{\mathbb R}}
\newcommand{\PP}{{\mathbb P}}
\newcommand{\CC}{{\mathbb C}}
\newcommand{\ZZ}{{\mathbb Z}}

\renewcommand{\Re}{\textrm{Re}\,}
\DeclareMathOperator{\sech}{sech}

\def\D{{\mathcal D}}

\renewcommand{\th}{\theta}

\newcommand{\e}{{\rm e}}

\newtheorem{thm}{Theorem}
\newtheorem{remark}{Remark}

\DeclareMathOperator{\Diff}{Diff}
\newcommand{\ip}[2]{\left<#1\mid#2\right>}

\begin{document}
\title{Probability distributions for the stress tensor in conformal field theories}
\author[1]{Christopher J. Fewster\thanks{\tt chris.fewster@york.ac.uk}}
\author[2]{Stefan Hollands\thanks{\tt stefan.hollands@uni-leipzig.de}}
\affil[1]{Department of Mathematics,
University of York, Heslington, York YO10 5DD, United Kingdom.}
\affil[2]{Institute for Theoretical Physics, University of Leipzig,
	Br\"{u}derstra{\ss}e 16,
	D-04103 Leipzig, Germany.} 
\date{2 August 2018}
\maketitle 

\begin{abstract}
The vacuum state -- or any other state of finite energy -- is not an eigenstate of any smeared (averaged) local quantum field. The outcomes (spectral values) of
repeated measurements of that averaged local quantum field are therefore distributed according to a non-trivial probability distribution.
In this paper, we study probability distributions for the smeared stress tensor in two dimensional conformal quantum field theory. 
We first provide a new general method for this task based on the famous conformal welding problem in complex analysis. 
Secondly, we extend the known moment generating function method of Fewster, Ford and Roman. Our analysis provides new explicit probability distributions for the smeared 
stress tensor in the vacuum for various infinite classes of smearing functions. All of these turn out to be given in the end by a shifted Gamma distribution, pointing, perhaps, at a distinguished role of this distribution in the problem at hand.
\end{abstract}
\bigskip
\noindent {\small Keywords: conformal field theory, conformal welding, moment problems, probability distributions, quantum energy inequalities.}


\section{Introduction}

According to the standard postulates of Quantum Theory, if an observable (self-adjoint operator) $A$ is measured repeatedly in  
a state $|\Psi\rangle$, then the possible measurement outcomes (spectral values) $\lambda$ of $A$
will be distributed according to a probability distribution. In fact, assuming for simplicity that the 
spectrum is discrete, the probability to measure the $n$-th (non-degenerate) eigenvalue $\lambda_n$ of $A$ is $p_n = |\langle \Phi_n | \Psi \rangle |^2$, 
where $|\Phi_n \rangle$ is the corresponding normalized eigenvector. The general case is covered by the spectral theorem, which provides a probability 
measure $\d \nu_{A,\Psi}(\lambda)$ given in terms of 
the spectral decomposition $P_A(\d \lambda)$ of $A$ by $\d \nu_{A,\Psi}(\lambda) = \ip{\Psi}{ P_A(\d \lambda) \Psi}$.\footnote{In the case of a self-adjoint matrix $A$, we have $P_A(\d \lambda) = \sum_n \delta(\lambda_n - \lambda) \d \lambda |\Phi_n \rangle \langle \Phi_n |$ and consequently 
$\d \nu_{A,\Psi}(\lambda) = \sum_n p_n \delta(\lambda -\lambda_n)\d\lambda$, i.e. 
the measure is singular w.r.t. the Lebesgue measure.} If $|\Psi \rangle$ is an eigenstate of $A$ the measure is concentrated solely on the corresponding eigenvalue; if it is not, then the probability distribution is non-trivial.

In Quantum Field Theory (QFT), observables of interest are the ``averaged'' local quantum fields, 
\ben\label{smeared}
A= \phi(f) \equiv \int \phi(x) f(x) \d^d x
\een
where $f$ is a smooth spacetime ``sampling function'', e.g. a smooth function that is non-zero within some bounded region. Of particular interest is the 
stress tensor of the theory, $\phi=\Theta_{\mu \nu}$, in which case \eqref{smeared} has the interpretation of 
an averaged stress-energy-momentum component within the spacetime region characterized by the sampling function $f$.  
A surprising feature of QFT which follows directly from the Reeh--Schlieder theorem~\cite{ReehSchlieder:1961,Haag} 
is that the vacuum state (or for that matter, any state with finite or exponentially bounded energy) 
is {\em never} an eigenstate of {\em any} $\phi(f)$ -- unless the field is trivial, i.e. a multiple of the identity operator. 

Thus, in quantum field theory, any state with finite or exponentially bounded energy -- such as e.g. an $N$-particle state -- 
gives a non-trivial probability distribution for any averaged local field $\phi(f)$.
If $\phi(f)$ has vanishing vacuum expectation value then there must be non-zero probabilities of obtaining both positive and negative outcomes from measurements $\phi(f)$ in the vacuum state; this is true even in case $\phi$ corresponds to a classical field observable (such as e.g. the energy density component of the stress-tensor) that is classically non-negative~\cite{EGJ}. What is more, the distribution is typically
strongly skewed, with large probabilities for small negative values balancing against small probabilities of large positive outcomes to give
an overall expectation value of zero in the vacuum state. 

Unfortunately, calculating the probability distribution of the smeared stress tensor seems a hard problem already in the simplest cases, e.g.
when the state is the vacuum and the underlying QFT describes free fields or conformal QFTs (CFTs) in $1+1$ dimensions. In fact, 
closed form analytical expressions seem to be available only in the last case so far.\footnote{
A recent series of papers by one of us, together with Ford, Roman and Schiappacasse 
has explored the probability distribution of measurements of energy densities and related observables in various situations~\cite{FewsterFordRoman:2010,FewForRom:2012,FewFord:2015,SchiaFewsFord:2018}. See Section~\ref{sec:conclusion} for some discussion.} 
There it was shown~\cite{FewsterFordRoman:2010} that Gaussian averages of the energy density in the vacuum state are distributed according
to a shifted Gamma distribution, whose parameters are given explicitly in terms
of the central charge $c$ of the theory and the variance of the Gaussian concerned. 
Being more specific, the averaged ``chiral half'' of the energy density operator is
\begin{equation}
\Theta(f)= \int_{-\infty}^\infty \Theta(u)f(u)\,\d u,
\end{equation}
where $\Theta(u)$ is the energy density operator on a light-ray (i.e., more properly speaking, the generator of translations in a light-like direction, see Section~\ref{sec:CFT}). Let $f(u)=\e^{-(u/\tau)^2}/ (\tau\sqrt{\pi})$
and consider measurements of $\Theta(f)$ in the vacuum state. The results are
statistically distributed according to a probability distribution that will be called the \emph{vacuum distribution of $\Theta(f)$}. The moments of this 
distribution can be organised into generating functions that are constrained
by conformal Ward identities. At least in principle this permits the 
moment generating function and probability distribution to be obtained in closed form.  In \cite{FewsterFordRoman:2010} it was shown (for Gaussian $f$) that the 
vacuum distribution has the {\em shifted Gamma  
probability density} 
\begin{equation}\label{eq:shiftedGammapdf}
\d \nu_f(\lambda) = \rho(\lambda; \alpha, \beta, \sigma) \d \lambda := 
\vartheta(\lambda+\sigma)\frac{\beta^\alpha(\lambda+\sigma)^{\alpha-1}}{\Gamma(\alpha)}  \exp(-\beta(\lambda+\sigma)) \, \d \lambda,
\end{equation}
where $\vartheta$ is the Heaviside function and the parameters $\alpha$, $\beta$ and $\sigma$  are
\begin{equation}\label{eq:Gaussian_pars}
\alpha=\frac{c}{24}, \qquad \beta=\pi\tau^2, \qquad \sigma = \frac{c}{24\pi\tau^2}. 
\end{equation}
Evidently, the vacuum distribution is supported in the half-line $[-\sigma,\infty)$;
on general grounds~\cite{FewsterFordRoman:2010}, the infimum of the support is equal to the optimal quantum energy inequality (QEI) for $\Theta(f)$,
\begin{equation}
-\sigma = \inf_\Psi \ip{\Psi}{\Theta(f)\Psi},
\end{equation}
with the infimum taken over all physically acceptable\footnote{In mathematical terms, 
we can for instance consider all states from the dense subset $\bigcap_{k \ge 0} \D (L_0^k)$, see sec. \ref{sec:CFT}, which is a core for 
$\Theta(f)$.} normalized states $\Psi$.
For unitary, positive energy CFTs, the optimal QEI was established rigorously in~\cite{Fe&Ho05},
building on an earlier argument of Flanagan~\cite{Flanagan:1997} for massless scalar fields. The bound is  
\begin{equation}
\label{QEI}
\inf_\Psi \ip{\Psi}{\Theta(f)\Psi}= -\frac{c}{12\pi}\int_{-\infty}^\infty \left(\frac{\d}{\d u}\sqrt{f(u)}\right)^2\,\d u  
\end{equation}
and reproduces the value $-\sigma$ when $f$ is the Gaussian. 

As mentioned above, this is the only known example of a sampling function $f$ for which the vacuum probability distribution has been determined in closed form. While the method of~\cite{FewsterFordRoman:2010} applies to general test functions, it involves the solution of a nonlinear differentio-integral flow equation as an intermediate step and the only solutions known until now are Gaussian in form. There is another (partly formal) approach, due to Baumann~\cite{Baumann}, that gives the characteristic function of the CFT vacuum probability distribution for general sampling functions in terms of a functional calculus expression. To our knowledge this has not been used to compute a closed form for the probability distribution in any specific example.

This paper makes progress in two directions. In part 1, we develop a novel method for computing probability distribution for the chiral smeared stress tensor in CFTs based on conformal welding. It works also for states other than the vacuum. In part 2 we develop further the moment generating technique of \cite{FewsterFordRoman:2010} by presenting new solutions to the flow equation, thereby obtaining new explicit formulas for probability distributions for sampling functions $f$ belonging to two infinite families (that then can be combined and manipulated in various ways). In addition, we extend the ideas of~\cite{FewsterFordRoman:2010} to thermal states.

\subsection{Part 1}

The idea behind our new method is to explore the well-known relationship between the operator $e^{i\Theta(f)}$ for smooth, compactly supported 
and real-valued functions $f$, and diffeomorphisms of the real line. In fact, the unitaries $e^{it\Theta(f)}, t \in \RR$ represent (up to phases) the action of the 1-parameter group of diffeomorphisms, $\rho_t$, flowing points $u$ of the real line along the vector field $f(u) \d / \d u$. We are able to convert the problem of calculating the characteristic function of the vacuum probability distribution $\d \nu_f(\lambda)$, 
\ben\label{chf}
\langle \Omega | e^{it\Theta(f)} \Omega \rangle=\int e^{it\lambda} \d \nu_f(\lambda) ,
\een
to a ``conformal welding problem'' along the diffeomorphisms $\rho_t$. By a conformal welding problem, one means in the simplest case the problem of 
finding a pair of univalent analytic ``welding'' maps $w^\pm$ from the upper/lower complex half plane $\HH^\pm$ to $\CC$ such that $w^+(u) = w^-(\rho(u))$ for all points $u$ on the real axis, where $\rho$ is some given diffeomorphism of the real line. Given a solution to this problem for the diffeomorphism $\rho_t(u)$ generated by $f(u)$, we show how to obtain from $w^\pm_t$ a solution to the problem of finding the probability distribution for $\Theta(f)$ in the vacuum. 

The method is conceptually interesting because it establishes a connection between CFT and the beautiful mathematical theory of such weldings, which is by now a classic part of complex analysis, see e.g. \cite{sharon} which also treats numerical implementations for solving the welding problem as well as a 
connection with 2d-shape recognition theory. We give a concrete example 
in which the welding maps and the probability distribution can be calculated analytically in closed form taking the Fourier transform of \eqref{chf}, yielding a 
generalized hyperbolic secant distribution. Furthermore, we show how to generalize the method to other states such as thermal or highest weight states in CFT.   

\subsection{Part 2}

In the second part, we develop further the moment generating method of~\cite{FewsterFordRoman:2010}. 
In this method, the probability distribution for any local average of the energy density in the vacuum state 
is expressed in terms of the solution to a certain nonlinear integro-differential flow equation whose initial condition is given by the averaging function $f$. 
We generalize this method to the case of a thermal (Gibbs-) state at some finite temperature. Instead of one nonlinear integro-differential flow equation, 
we now get a coupled system, whose initial condition is given by the averaging function $f$ and the inverse temperature $\beta$. The 
method relies on the Ward-identities for the stress tensor in a thermal state \cite{felder}, in a similar way as the method of~\cite{FewsterFordRoman:2010}
relied on those in the vacuum state.

Even in the vacuum, the flow equation was solved in~\cite{FewsterFordRoman:2010} only for Gaussian $f$. The main novelty in part 2 is that we are able to present two infinite new families of solutions to this equation. These give rise to two new infinite families
of averaging functions $f$ for which the vacuum distribution can be obtained in 
closed form. They are given by powers of the Lorentzian function, and  
a family related to the inverse Gamma distribution. In these cases, the 
vacuum distribution turns out to be a shifted Gamma distribution, just as in the
case where $f$ is Gaussian. It is noteworthy that the 
Gaussian, Lorentzian and inverse Gamma are examples of \emph{stable distributions}: that is, the sum of independent random variables distributed according to such a distribution is a member of the same family. We conjecture that all stable distributions can be analysed in a similar way.

\medskip
\noindent
The structure of this paper is as follows. In sec. \ref{sec:CFT} we first review some well-known CFT basics. We then describe our general 
method based on conformal welding in sec. \ref{sec:CW}, and then we describe our results based on the moment generating function technique in 
sec. \ref{sec:MGF}. The evaluation of some integrals is moved to an appendix. 

\section{Notation and CFT basics}\label{sec:CFT}

Here we describe our notation and basic facts about the stress tensor in two-dimensional conformal field theories (CFTs). Our conventions follow those used in~\cite{Fe&Ho05}; for a more detailed exposition of CFT in particular in relation with vertex operator algebras, see \cite{carpi}. As is well-known, the stress energy operator in a CFT on $(1+1)$-dimensional Minkowski spacetime has two independent, commuting (``left and right chiral'') components depending only on the left and right moving light-ray coordinates $u=x^0-x^1, v=x^0+x^1$, respectively. Focussing on one of them, we get 
a quantum field $\Theta(u)$ living on one of the light-rays. A light-ray may be compactified to a circle via the Cayley transform, and in this way we get a 
quantum field $T(z)$ on the circle. In order to set up the theory  in a mathematically precise way, it is in some sense most natural 
to turn this story around and start from the quantum field $T(z)$ on the circle, which we shall do now. 

The basic algebraic input is the Virasoro algebra. It is the Lie-algebra with generators  $\{ L_n, \kappa \}_{n \in \ZZ}$  obeying
\ben
[L_n,L_m] = (n-m) L_{n+m} + \frac{\kappa}{12} n(n^2-1) \delta_{n,-m} , \quad 
[L_n, \kappa] = 0 .
\een  
A positive energy representation on a Hilbert space $\H$ is a representation such that (i) $L^*_n=L_{-n}$ (unitarity), (ii) $L_0$ is diagonalizable with 
non-negative eigenvalues, and (iii) the central element is represented by $\kappa = c \myid$. 
From now, we assume a positive energy representation. We assume that $\H$ 
contains a vacuum vector $|\Omega\rangle$ which is annihilated by $L_{-1}, L_0, L_1$, 
($\mathfrak{sl}(2,\RR)$-invariance) and which is a highest weight vector (of weight 0), i.e. $L_n |\Omega\rangle = 0$ for all 
$n >0$. One has the bound \cite{carpi,buchholz,goodman_wallach,goodman_wallach1}
\ben
\label{poly}
\|(1+L_0)^k L_n \Psi \| \le \sqrt{c/2}(|n|+1)^{k+3/2} \|(1+L_0)^{k+1} \Psi \|
\een
for $|\Psi \rangle \in \bigcap_{k \ge 0} \D (L_0^k) \subset \H$ and any natural number $k$.

One next defines from the Virasoro algebra the stress tensor on the unit circle $\SS$, identified with points $z=e^{i\theta}, \theta \in \RR$ in $\CC$. The stress tensor is an operator valued distribution on $\H$ defined in the sense of distributions by the series
\ben
T(z) = -\frac{1}{2\pi} \sum_{n =-\infty}^\infty L_n z^{-n-2}. 
\een
More precisely, for a test function $f \in C^\infty(\SS)$ on the circle, 
it follows from \eqref{poly} that the corresponding smeared field 
\ben
T(f) = \int_{\SS} T(z) f(z) \d z :=-\frac{1}{2\pi} \sum_{n =-\infty}^\infty  \left(\int_{\SS} z^{-n-2} f(z) \d z\right)L_n
\een
is an operator defined e.g. on the dense invariant domain $\bigcap_{k \ge 0} \D (L_0^k) \subset \H$ (which can be shown to 
be a common core for the operators $T(f)$) and the assignment 
$f \mapsto T(f)|\psi \rangle$ is continuous in the topologies on $C^\infty(\SS)$ and $\H$ for any vector in this domain.
Letting $\Gamma$ be the anti-linear involution 
\begin{equation}
\Gamma f(z) = -z^2 \overline{f(z)},
\end{equation} 
the smeared stress tensor is a self-adjoint 
operator on $\D(L_0)$ for $f$ obeying the reality condition $\Gamma f = f$, and  one has $T(f)^* = T(\Gamma f)$ in general. 

A real test function (in the above sense) defines a real vector field ${\sf f}
\in {\rm Vect}_\RR(\SS)$ by means of the formula 
\ben\label{ff}
({\sf f} g)(z) = f(z) g'(z), 
\een
where $ie^{i\theta}g'(e^{i\theta}) = \frac{\d}{\d \theta} g(e^{i\theta})$. Under this correspondence, if we 
define $l_n(z) = z^{n+1}$ then the corresponding complex vector fields ${\sf l}_n = z^{n+1} \frac{\d}{\d z} \in {\rm Vect}_\CC(\SS) = {\rm Vect}_\RR(\SS) \otimes_\RR \CC$ satisfy the Witt algebra
\ben
[{\sf l}_n, {\sf l}_m]=(m-n) {\sf l}_{n+m}
\een
under the commutator of vector fields, and furthermore $iT(l_n) = L_n$.  

For real $f \in C^\infty(\SS)$, we denote by $\rho_t \in {\rm Diff}(\SS)$ the 1-parameter flow of diffeomorphisms generated by 
the corresponding vector field ${\sf f}$, in formulas, 
\ben\label{flow}
\frac{\partial}{\partial t} \rho_t(z) = f(\rho_t(z)), \quad \rho_0 = \rm{id}. 
\een 
Note that $\rho_t$ leaves invariant all $z$ outside the support of $f$. 
For a smooth function $f(z)$ on 
the complex plane or circle, the Schwarzian derivative is defined by 
\ben
Sf(z) = \frac{f'''(z)}{f'(z)} - \frac{3}{2} \left( \frac{f''(z)}{f'(z)} \right)^2 . 
\een
It can be shown (see~\cite{Fe&Ho05}, which uses results of \cite{goodman_wallach,goodman_wallach1,toledano})
that $e^{iT(f)}$ (for real $f$) leaves invariant the dense set $\bigcap_{k \ge 0} \D (L_0^k) \subset \H$ of vectors, and on this set, 
we have the transformation formula
\ben\label{rel1}
e^{iT(f)} T(z) e^{-iT(f)} = \rho'(z)^2 T(\rho(z)) - \frac{c}{24 \pi} S\rho (z) \, \myid , 
\een
in the sense of distributions in the variable $z \in \SS$, 
where $\rho = \rho_{t=1}$ is the flow of $f$ at unit flow-`time', i.e., the exponential $\rho=\exp{\sf f}$. See \cite{Fe&Ho05} for the somewhat non-trivial 
assignment of phases and the unitary implementation of the covering group of  $\Diff_+(\SS)$, which corrects an error in \cite{goodman_wallach}. 

 The generators $L_0, L_{\pm 1}$ together exponentiate to a unitary positive energy representation of  $\rm{PSL}(2,\RR)$ on $\H$ leaving $|\Omega\rangle$ invariant. Under this action the stress tensor transforms covariantly as in \eqref{rel1} with action $z \mapsto \rho(z) = (az+b)/(\bar b z + \bar a)$ on the circle, where the correspondence between the matrix $
 \left(
 \begin{matrix}
 a & b\\
 \bar b & \bar a
 \end{matrix}
 \right) \in \rm{SU}(1,1), 
 $  and a group element of $\rm{PSL}(2,\RR)$ 
 is given by the standard group isomorphism $\text{SU}(1,1) \cong \rm{PSL}(2,\RR)$.

The stress tensor on the real line is defined by pulling back the stress tensor on the circle via the Cayley transform $C$, which maps
the real line to the circle by 
\begin{equation}
\RR \owns u \mapsto C(u)=\frac{1+iu}{1-iu} \in \SS \setminus \{-1\},
\end{equation} 
with inverse $C^{-1}(z)=i(1-z)/(1+z)$. Defining $C(\infty)=-1$ this becomes a bijection between the compactified real line and the circle.
Then the stress tensor on the real line is
\ben\label{real}
\Theta(u) \equiv  \left( \frac{\d C(u)}{\d u} \right)^2 T(C(u)) = -\frac{4}{(1-iu)^4} T\left(\frac{1+iu}{1-iu} \right), 
\een 
in the sense of an operator valued distribution on the same domain. Using this formula, we can easily convert any result on stress tensor on 
the circle to one on the real line. 

Finally, the stress energy tensor in a $1+1$-dimensional Minkowski spacetime $\RR^{1,1}$ is constructed on the tensor product $\H \otimes \H$ carrying two 
(not necessarily equal) positive energy representations of the Virasoro-algebra. These allow us to define $\Theta^R(u) = \Theta(u) \otimes 1, \Theta^L(v) = 
1 \otimes \Theta(v)$, from which the components of $\Theta_{\mu\nu}(x)$ are then defined as 
Wightman fields on $\RR^{1,1}$ in terms of inertial coordinates $x=(x^0,x^1)=\frac12(u+v,-u+v)$ by
\ben
(\Theta^{\mu\nu}(x)) = 
\left(
\begin{matrix}
\Theta^R(u) + \Theta^L(v) & \Theta^R(u) - \Theta^L(v)\\
\Theta^R(u) - \Theta^L(v) & \Theta^R(u) + \Theta^L(v)
\end{matrix}
\right).
\een
The stress energy tensor satisfies $\partial^\mu \Theta_{\mu\nu} = 0 = \Theta^\mu_\mu$. 
Conversely, the L\" uscher-Mack theorem \cite{luscher} states that any translation and dilation invariant hermitian Wightman field theory containing an operator valued distribution
$\Theta_{\mu\nu}$ such that $P_\nu=\int \Theta_{\mu\nu}(x) \d x^\mu$ (with the integral over any $x^0=cst.$ surface) is the generator of translations, the 
left- and right moving components $\Theta^{L/R}$ satisfy the relations of two commuting Virasoro algebras. 
In particular, if we have probability distributions of $\Theta^{L/R}(f)$ in some state, we can therefore immediately find the corresponding probability distribution of any component of $\Theta_{\mu\nu}$.

\section{Part 1: Probability distributions from conformal welding}\label{sec:CW}

\subsection{General construction}

Our goal is to describe a general method for obtaining, in principle, the probability distribution of $T(f)=\int T(z) f(z) \d z$ in a vacuum state $|\Omega\rangle$. This 
method will be based on the technique of conformal welding. We assume that $f \in C^\infty(\SS)$ is a test function on the circle satisfying the reality condition $\Gamma f = f$ which has its support within some interval $I \subset \SS$. Since $T(f)$ is self-adjoint, the probability distribution is a measure $\d \nu_f(\lambda)$ on $\RR$ in view of the spectral theorem. It is uniquely determined by its characteristic function
\ben
\int_\RR e^{it\lambda} \d \nu_f(\lambda) = \langle e^{itT(f)} \rangle , 
\een
where $t \in \RR$ and where and, as in the rest of this subsection, $\langle A \rangle = \langle \Omega| A\Omega \rangle$. 
Our aim is to find this characteristic function in terms of $f$. 

To this end, we define an auxiliary function $G(z)$ by
\ben
G(z) = \langle e^{iT(f)} \rangle^{-1} 
\begin{cases}
\langle T(z) e^{iT(f)} \rangle & \text{if $|z|>1$,}\\
\langle e^{iT(f)} T(z) \rangle & \text{if $|z|<1$.}
\end{cases}
\een 
A priori, both functions $\langle T(z) e^{iT(f)} \rangle$ and  $\langle e^{iT(f)} T(z) \rangle$ are only defined as a distribution on $\SS$. However,
in view of \eqref{poly},  the series defining $T(z) |\Omega \rangle$ converges absolutely for $|z| < 1$, while the series defining $\langle \Omega | T(z)$
converges absolutely for $|z|>1$. The properties $L_n |\Omega\rangle = 0$ for $n \ge -1$ then show that $G(z)$ is a holomorphic function on $\CC \setminus \SS$ vanishing at infinity as $z^{-4}$. 

Now let $\DD^+=\{|z|<1\}$ be the interior of the circle $\SS$ and $\DD^-=\{|z|>1\}$ the exterior. Using the energy bound \eqref{poly} it follows that the 
limits from inside/outside the disk 
\ben
G^\pm(z) = \lim_{r \to 1^\mp} G(rz)
\een
for $z \in \SS$ exist in the sense of distributions on $\SS$. For $z \in \SS \setminus I$ in the complement of the interval $I$ where $f$ is supported, the 
relation \eqref{rel1}  gives immediately
\ben
G^+(z) = \frac{\langle  e^{iT(f)} T(z) \rangle}{\langle e^{iT(f)} \rangle} = \frac{\langle T(z) e^{iT(f)}   \rangle}{\langle e^{iT(f)} \rangle} = G^-(z), 
\quad 
\text{if $z \in \SS \setminus I$.}
\een
On the other hand, for $z \in I$, we get using again \eqref{rel1}, 
\ben\label{jump1}
\begin{split}
G^+(z) &= \frac{\langle  e^{iT(f)} T(z) \rangle}{\langle e^{iT(f)} \rangle} \\
&= \frac{\langle \{ \rho'(z)^2 T(\rho(z)) - \frac{c}{24 \pi} S\rho (z) \myid \} e^{iT(f)}  \rangle}{\langle e^{iT(f)} \rangle} \\
&= \rho'(z)^2 G^-(\rho(z)) - \frac{c}{24 \pi} S\rho (z), 
\quad 
\text{if $z \in I$.}
\end{split}
\een
By the edge-of-the-wedge theorem $G(z)$ is hence a holomorphic function on $\CC \setminus I$. It vanishes at infinity as $O(z^{-4})$, and satisfies across $I$ the jump condition \eqref{jump1}. We shall now argue that it is possible to reconstruct $G$ from this information. 

We first consider two univalent holomorphic functions $w^\pm: \DD^\pm \to \Delta^\pm$ from the inside ($+$) or outside ($-$) of the circle onto the inside/outside $\Delta^\pm$ of a Jordan curve $C$, such that their respective boundary values on the circle $\SS$ satisfy the junction (``conformal welding'') condition
\ben\label{jump2}
w^+(z) = w^- \circ \rho(z) \quad \text{for $z \in \SS$.}
\een
The existence of such functions, given the diffeomorphism $\rho$, is a classic result, see e.g. \cite{sharon} and references therein; we will recall two methods for constructing the solution below. $w^\pm$ may be normalized in such a way that the point $z=0$ gets mapped to zero with unit differential at $0$, and that $z=\infty$ gets mapped to infinity with unit differential on the real axis at infinity. The functions $w^\pm$ can be joined together to a holomorphic map defined on $\CC \setminus I$ called $w$, which is invertible on its image, with an inverse $z(w)$. Note that the normalization conditions on $w_+$ imply that $w''$ and $w'''$ vanish as $z\to\infty$ so $(Sw)(z)$ and thus $(Sz)(w)$ also vanish at infinity. Consider next the function $H$ defined on $\CC$ minus the portion of 
the Jordan curve $C$ which is the image of $I$, 
\ben\label{Hdef}
H(w) = z'(w)^2 G(z(w)) - \frac{c}{24 \pi} Sz (w). 
\een
Also, define the boundary values from the inside resp. outside of the Jordan curve $C$ as 
\ben\label{Hdef1}
H^\pm(w) = \lim_{v \to w, v \in \Delta^\pm} H(v). 
\een
Using the chain rule for the Schwarzian derivative, 
\ben
S(f \circ g)(z) = g'(z)^2 (Sf) \circ g(z) + Sg(z) ,  
\een
as well as the jump conditions \eqref{jump1}, \eqref{jump2}, we immediately see that 
\ben
H^+(w) = H^-(w) \quad \text{for $w \in C$,}
\een
and from the normalization of $w^\pm$, we see see that $H(w)$ is holomorphic on $\Delta^\pm$ and vanishes at infinity. By the edge-of-the-wedge theorem, $H(w)$ is hence holomorphic on all of $\CC$ and vanishes at infinity, and hence must be zero. We conclude that $G(z) = 
\frac{c}{24 \pi} (z' \circ w)^{-2} (S\rho) \circ w$, and thus in view of the chain rule for the Schwarzian derivative that 
\ben
G(z) = -\frac{c}{24 \pi} Sw(z)
\een
on $\CC \setminus I$. 
We can obviously repeat the same construction with the $\Gamma$-invariant test-function $tf(z), t \in \RR$, resulting in a $t$-dependent function $w(t,z)$ and 
a $t$-dependent function $G(t,z)$. 
By construction, the characteristic function of the probability measure then satisfies, for either $+$ or $-$,  
\ben
\label{result0}
\frac{\partial}{\partial t} 
\log \int_\RR e^{it\lambda} \d \nu(\lambda) = \frac{\partial}{\partial t} \log \langle e^{itT(f)} \rangle 
= i \int_{\SS} G^\pm(t,z) f(z) \d z , 
\een
or integrating 
\ben
\label{result1}
\langle e^{iT(f)} \rangle = \exp \left( -\frac{ic}{24 \pi} \int_0^1 \int_{\SS} f(z) Sw^-(t,z) \, \d z \d t \right) , 
\een
which would also hold for  $+$ instead of $-$. Thus, we have in principle determined the characteristic function of the probability measure: We must first solve the 
flow equation \eqref{flow} to determine $\rho(t, z)$ and then the jump problem \eqref{jump2} to determine $w(t,z)$. 

\begin{remark} If we integrate \eqref{result0} between $0$ and $t$, we can also write the result as
\ben\label{result}
\int_\RR e^{it\lambda} \d \nu_f(\lambda) = \exp \left( -\frac{ic}{24 \pi} \int_0^t \int_{\SS} f(z) Sw^-(t',z) \, \d z \d t' \right) , 
\een
\end{remark}

For completeness, we now recall two methods to solve the jump problem \eqref{jump2}, setting again $t=1$ and $\rho(z) = \rho_1(z)$ to simplify the notation. Both methods can, in principle, be implemented numerically, see e.g. \cite{sharon}. 
Since $\rho$ is in the connected component of the identity in ${\rm Diff}(\SS)$, there is a smooth function $\chi: \RR \to \RR$ with 
$\chi(\theta + 2\pi) = \chi(\theta) + 2\pi$ and with $\chi'(\theta)>0$ such that $\rho(e^{i\theta}) = \exp i\chi(\theta)$. With the aid of this function, we extrapolate the diffeomorphism 
$\rho \in {\rm Diff}(\SS)$ to a homeomorphism of the unit disk $\DD^+$ as
$
\rho(re^{i\theta}) = re^{i\chi(\theta)}. 
$
Now let 
\ben
\mu(z) = 
\begin{cases}
\partial_{\bar z} \rho(z)/\partial_{z} \rho(z) & \text{if $z \in \DD^+$,}\\
0 & \text{otherwise.}
\end{cases}
\een
This can also be written as $\mu(re^{i\theta}) = e^{i\theta} (1-\chi'(\theta))/(1+\chi'(\theta))$ on $\DD^+$ and $\mu(z)=0$ on $\DD^-$. 
In view of $\chi'(\theta)>0$, it follows that $\mu \in \L^\infty(\CC)$, and in fact
${\rm ess-sup} |\mu| < 1$. It is well-known (see e.g. \cite{ivaniec} for a detailed exposition) 
that the Beltrami equation 
\ben
\partial_{\bar z} F = \mu \partial_z F 
\een 
has a ``principal solution'' under this condition. This solution is a homeomorphism of the inside/outside of the unit disk
to the inside/outside of a Jordan curve $C$ and we may apply a linear transformation replacing $F$ by $aF+b$ with $a,b \in \CC$ to achieve that $F(0)=0$ and $F$ has unit derivative at the point at infinity on the real axis. Note that $\rho$ satisfies the Beltrami equation on $\DD^+$, so by the Stoilov factorization theorem (see e.g. \cite{ivaniec}), it must be the case that 
$F$ is $\rho$ followed by an analytic function on $\DD^+$. So let $w^+$ be the analytic function $F \circ \rho^{-1}$ on $\DD^+$ and let $w^-=F$ on $\DD^-$. It follows that 
$w^\pm$ have the desired properties.

One may also obtain a more explicit expression for $w^\pm$ which is manifestly independent of the precise way of extending $\rho$ to the disk via the Hilbert transform, see \cite{gakhov} for details. First, $w^\pm(z)$ are assumed to be given in $\DD^\pm$ by convergent power series
\ben
w^+(z) = z + a_2 z^2 + a_3 z^3 + \dots, \quad w^-(z) = z + b_1 z^{-1} + b_2 z^{-2} + \dots . 
\een
On the Hilbert space $\L^2(\SS) \equiv \L^2(\SS, \d \theta/2\pi)$ consider the orthonormal basis $e_n(\theta) = e^{in\theta}$ where $n \in \ZZ$. 
It follows that  $w^+$ is in the closed subspace $\H^2(\SS)$ spanned by $e_n, n \ge 0$, which can be identified with the Hardy space of 
holomorphic functions on $\DD^+$ that are square integrable on each 
circle $\SS_r=\{ re^{i\theta} \}, r<1$ with uniformly bounded $\L^2(\SS_r)$-norm.  Let $\Sigma$ be the Hilbert transform defined by 
$\Sigma e_n = {\rm sgn}(n) e_n$ for $n \neq 0$ and $\Sigma e_0=0$. The integral kernel of $\Sigma$ with respect to the measure $\d \theta/(2\pi)$ is 
$\Sigma(\theta, \theta')=i \ctn(\frac{\theta-\theta'}{2})$ for $\theta \neq \theta'$. Viewing $w^-$ as a function on $\SS$, 
we have $\Sigma w^- = -w^-  + 2e^{i\theta}$ and from the jump condition \eqref{jump2} we also have $\Sigma (w^- \circ \rho) = w^- \circ \rho$. It follows 
$\Sigma w^- -\Sigma(w^- \circ \rho) \circ \rho^{-1}  = -2w^- +2e^{i\theta}$. Let $U_\rho F(z) = F(\rho^{-1}(z))$. Then $U_\rho$ is a bounded operator on $\L^2(\SS)$, 
and, letting $K=(\Sigma-U_\rho \Sigma U_\rho^{-1})/2$, we have $(I+K)w^- =e^{i\theta}$. Explicitly, for a smooth function $F$ such that $F(\theta +2\pi) = F(\theta) + 2\pi$ the integral operator $K$ is
\ben\label{Kdef}
\begin{split}
KF(\theta) &=
\frac{i}{4\pi} \int_0^{2\pi} \left[ \ctn \left( 
\frac{\theta-\theta'}{2}
\right) -  (\chi^{-1})'(\theta') \, \ctn \left( 
\frac{\chi^{-1}(\theta)-\chi^{-1}(\theta')}{2}
\right)
\right] F(\theta') \d \theta', 
\end{split}
\een
where as before, $\rho(e^{i\theta})= e^{i\chi(\theta)}$. It can be shown that the kernel of this operator is smooth, and that $I+K$ has no kernel. By the Fredholm alternative, it is invertible, and therefore
\ben\label{wdef}
w^-(e^{i\theta}) = (I+K)^{-1}(e^{i\theta}) . 
\een
So our main result is as follows.

\begin{thm}
Let $f(z)$ be a $\Gamma$-invariant smooth function on $\SS$, let $\chi_t(\theta)$ be the flow defined by $\partial_t \chi_t(\theta) = -ie^{-i\chi_t(\theta)}f(e^{i\chi_t(\theta)})$, let $w^-(t,z)$ be defined by \eqref{wdef} with $\chi_t$ in the definition of $K$, eq. \eqref{Kdef}. Then the characteristic function of the probability distribution 
of the self-adjoint operator $T(f)$ in the vacuum state $|\Omega \rangle$ is given by \eqref{result}.
\end{thm}

\begin{remark}
Our expression \eqref{result} for $\langle e^{iT(f)} \rangle$ is somewhat similar but not identical to an expression given by \cite{aldrovandi} and 
by \cite{Haba:1989in} for the exponential of the smeared stress tensor in {\em Euclidean} CFTs. In fact, their expression also involves the Schwarzian of 
a function $F(t,z)$ which is a solution to a Beltrami-equation. However, the Beltrami coefficient in their equation is not identical to our coefficient $\mu$, and 
their solution is consequently different. These differences are probably due to the fact that the exponential of the smeared stress tensor in Euclidean CFT would correspond more closely to a ``radially ordered'' exponential of a smeared stress tensor over $\CC$, whereas our expression concerns instead the non-ordered 
smeared stress tensor over $\SS$.  We believe that these differences are also at the root of an inconsistency between two derivations given in \cite{Haba:1989in} and noted by the author himself.
\end{remark}

\subsection{Generalizations}

\subsubsection{Light ray picture}
The expression \eqref{real} for the stress tensor on the light ray $\RR$ allows us to transplant the previous result to the light ray via the 
Cayley transform $u \owns \RR \mapsto C(u)=(1+iu)/(1-iu) \in \SS$. Let $g \in C^\infty_0(\RR, \RR)$ be a smooth real-valued test function
on the light ray and note that $\Theta(g)=T(f)$, where
\begin{equation}
f(z)=C'(C^{-1}(z))g(C^{-1}(z)).
\end{equation}
The method of the previous subsection, applied to $f$, therefore provides the characteristic function for $\Theta(g)$. It is instructive to see how the results can be expressed more directly in terms of $g$.
Let $\psi_t$ be the flow of the corresponding vector field ${\sf g} = g(u) \frac{\d}{\d u}$, i.e. 
\begin{equation}
\partial_t \psi_t(u) = g(\psi_t(u)), \qquad \psi_0 = \rm{id},
\end{equation}
and which is related to the flow $\rho_t$ induced by $f$ on $\SS$ by $\psi_t=C^{-1}\circ \rho_t\circ C$. Writing $w^\pm$ for the solution to the welding problem for $\rho=\rho_1$, we find that $w_\RR^\pm= C^{-1}\circ w^\pm\circ C$ 
solves the welding problem
for two holomorphic functions $w^\pm_\RR$ from the upper/lower half plane 
$\HH^\pm$ to $\CC$ leaving $\pm i$ fixed and such that their boundary values on the real axis satisfy 
\begin{equation}\label{eq:weldR}
w_\RR^+ = w_\RR^-\circ \psi
\end{equation} 
(here $\psi=\psi_1$). If we replace more generally $\psi$ by $\psi_t$, we get a solution $w^-_\RR(t,u)$ depending also on $t$.
Combining \eqref{real} with the fact that $SC(u)=0$, with the chain rule for the Schwarzian, and with the preceding theorem then gives us
\ben\label{result2}
\langle e^{i\Theta(g)} \rangle = \exp \left( -\frac{ic}{24 \pi} \int_0^1 \int_\RR g(u) Sw^-_\RR(t,u) \, \d u \d t \right) , 
\een
which is the analogue of the result \eqref{result1} for the light ray.

The welding problem may be solved by operator means, using the Cayley transform 
to define a unitary $U:\L^2(\SS,\d\theta/(2\pi))\to\L^2(\RR,\d u/(\pi(1+u^2)))$ 
by $UF=F\circ C$, the pull back. Then (compare \eqref{wdef})
\begin{equation}\label{wdef1}
\tilde{w}_\RR^- := (I+K_\RR)^{-1}f_1, \qquad K_\RR=UKU^{-1},\quad 
f_1(u)=(Ue_1)(u)=\frac{1+iu}{1-iu},
\end{equation}
is nonlinearly related to $w_\RR^-$ by $w_\RR^-=C^{-1}\circ \tilde{w}_\RR^-$.  In fact, one can use $\tilde{w}_\RR^-$ in place of $w_\RR^-$ in~\eqref{result2} because they have the same Schwarzian derivative due to the chain rule and $(SC)(u)=0$.

Explicitly, $K_\RR F = (K(F \circ C^{-1}))\circ C$ is the pull-back of the Fredholm operator $K$ \eqref{Kdef} to the light ray via the Cayley transform, and can be written
\begin{equation}
(K_\RR F)(u)=\frac{i}{2\pi}\int_{-\infty}^\infty \left[\frac{1}{u'^2+1}\frac{1+uu'}{u-u'}
-\frac{(\psi^{-1})'(u)}{1+(\psi^{-1})'(u')^2}\frac{1+\psi^{-1}(u)\psi^{-1}(u')}{\psi^{-1}(u)-\psi^{-1}(u')}\right]F(u')\d u'
\end{equation} 
for $\psi=\psi_1$. 
 
\subsubsection{KMS states}
By a similar trick, we can also obtain a corresponding result for KMS-states on the light ray. As is well-known, there is a  KMS state $\langle \ \cdot \ \rangle_\beta$ at each $\beta>0$ for the Virasoro algebra which is obtained by ``pulling back'' the vacuum state via the map $\RR \owns s \mapsto u(s)=e^{2 \pi s/\beta} \in (0,\infty)$ due to the Bisognano-Wichmann theorem in conformal field theory. This is in fact the unique KMS state by the results of \cite{camassaI,camassaII}. The expectation value of $e^{i\Theta(g)}$, $g \in C^\infty_0(\RR, \RR)$ in such a state is then given by 
\ben
\langle e^{i\Theta(g)} \rangle_\beta = \langle e^{i\Theta(g_\beta)} \rangle_\infty,
\een 
where the subscript $\infty$ has been inserted to indicate the vacuum state on the light ray, and where 
\ben\label{eq:gbeta}
g_\beta(u) = \frac{2 \pi u}{\beta} 1_{(0,\infty)}(u) \, g((\beta/2\pi) \log u) . 
\een 
By solving the welding problem~\eqref{eq:weldR} for the flow $\psi_t$ induced by $g_\beta$ to obtain $w_\RR^-$ the characteristic function for the thermal probability distribution is then given by~\eqref{result2}. The result can be expressed in terms of a welding problem for the $1$-parameter flow $\phi_t$ generated by the vector field
${\sf g} = g(s) \frac{\d}{\d s}$ on $\RR$ associated with $g$ and which is related
to $\psi_t$ by $\phi_t = u^{-1}\circ\psi_t\circ u$, i.e., 
\begin{equation}
\phi_t(s)=\frac{\beta}{2\pi}\log\psi_{t}(e^{2\pi s/\beta}).
\end{equation}
Consider the particular diffeomorphism $\phi=\phi_1$ and let $w_\RR^\pm$ solve the welding problem~\eqref{eq:weldR} for $\psi=\psi_1$. Defining $w_\beta^\pm(s)=\frac{1}{4}\beta w^\pm_\RR(e^{2\pi s/\beta})$, we obtain holomorphic maps from $\PP^\pm$ to $\CC$, where $\PP^\pm=\{z\in\CC:0<\pm\textrm{Im}\, z<\beta/2\}$ are the open 
upper/lower strips, leaving $\pm i\beta/4$ fixed and with boundary values on $\RR$ obeying
\begin{equation}
w_\beta^+(s)=w_\beta^-\circ\phi(s), \quad w_\beta^\pm(s \pm i\beta/2) = w_\beta^\pm(-s).
\end{equation}
Conversely, a solution to this welding problem for $\phi_t$ provides a solution to the welding problem~\eqref{eq:weldR} for $\psi_t$, whose Schwarzian derivative is easily found using the chain rule and  the fact that $S(e^{2\pi s/\beta}) = -2\pi^2 /\beta^2$. 
In this way, and using also the definition~\eqref{eq:gbeta}, there follows the formula
\ben
\langle e^{i\Theta(g)} \rangle_\beta = \exp \left( -\frac{ic}{24\pi} \int_0^1 \int_\RR g(s) Sw^-_\beta(t,s) \, \d s \d t \right) \exp \left( \frac{ic\pi}{12 \beta^2}
\int_\RR g(s) \, \d s \right) . 
\een

As before, $w_\beta^-$ may be obtained by Hilbert space means, though with a slight complication. Consider the diffeomorphisms $\phi=\phi_1$, $\psi=\psi_1$. Pull-back using $u(s)$ induces a partial isometry
\begin{equation}
V_\beta:\L^2(\RR,\d u/(\pi(1+u^2)))\to \L^2(\RR,\d s/(\beta \cosh(2\pi s/\beta)))
\end{equation} 
given by $(V_\beta F)(s)= F(u(s))=F(e^{2\pi s/\beta})$, whose adjoint $V_\beta^*$ is an isometry onto the subspace of functions supported in $\RR^+$. It is convenient to regard $\L^2(\RR,\d u/(\pi(1+u^2)))$ as the direct sum of the subspaces of functions supported on the positive and negative half-lines. In an obvious matrix notation, 
\begin{equation}
\begin{pmatrix} I+K_{\RR,++} & K_{\RR,+-} \\ K_{\RR,-+} & I+K_{\RR,--}\end{pmatrix}
\begin{pmatrix} w_{\RR,+}^- \\ w_{\RR,-}^-\end{pmatrix} = 
\begin{pmatrix} f_{1,+}  \\ f_{1,-} \end{pmatrix}
\end{equation}
where $f_{1,+}$, for instance, denotes the restriction of $f_{1,+}$ to the positive half-line. 
An important point is that the component $K_{\RR,--}$ of $K_\RR$ vanishes, because $\psi$ fixes all points on the negative half-line. This means that $w_{\RR,-}^-$ can be eliminated easily, leaving the equation
\begin{equation}
(I+K_{\RR,++} - K_{\RR,+-}K_{\RR,-+})w_{\RR,+}^- = f_{1,+}- K_{\RR,+-}f_{1,-}
\end{equation}
for $w_{\RR,+}$. As $I+K_\RR$ is invertible, the same is true of the operator on the left-hand side, giving altogether
\begin{equation}
w_{\RR,+}^- = (I+K_{\RR,++} - K_{\RR,+-}K_{\RR,-+})^{-1}(f_{1,+}- K_{\RR,+-}f_{1,-}).
\end{equation}
Restricted to the subspace corresponding to the positive half-line, $V_\beta$ becomes unitary, and we therefore have 
\begin{equation}\label{wdef2}
w_\beta^{-} = \frac{\beta}{4} (I+K_\beta)^{-1}g_{1,\beta} ,
\end{equation}
where $K_\beta = V_\beta (K_{\RR,++} - K_{\RR,+-}K_{\RR,-+}) V_\beta^*$ is compact
and $g_{1,\beta}=V_\beta (f_{1,+}- K_{\RR,+-}f_{1,-})$. 
An explicit kernel for $K_\beta$ can be written down, but we refrain from doing so in full. 

\subsubsection{Highest weight states}

We now assume that on the Hilbert space $\H$, we have operators $\phi_n, n \in \ZZ$ satisfying (i) an energy bound of the type \eqref{poly}, 
that is $ \|(1+L_0)^m \phi_n \Psi \| \le C(1+|n|)^s \| (1+L_0)^{k+m} \Psi \|$ for all $m \in {\mathbb N}_0, |\Psi \rangle \in \H$, and 
for some $s,k \ge 0$, satisfying (ii) the commutation relations 
$[L_m, \phi_n]=((h-1)m-n)\phi_{n+m}$, as well as (iii) $\phi_n |\Omega \rangle = 0$ for $n>-h$, where $h \in \mathbb N$. 
It follows from the energy bound (i) that for any test function $f \in C^\infty(\SS)$, the smeared ``primary field''
\ben
\phi(f) = \int_{\SS} \phi(z) f(z) \d z := \sum_{n =-\infty}^\infty  \left(\int_{\SS} z^{-n-h} f(z) \d z\right)\phi_n
\een
is an operator defined e.g. on the dense invariant domain $\bigcap_{k \ge 0} \D (L_0^k) \subset \H$, 
which is a common core for the operators $\phi(f)$. The properties (i) and (ii) imply furthermore that $\phi(z)| \Omega \rangle$ can be analytically continued to a vector-valued holomorphic function on $\DD^+$ with vector valued distributional boundary value on $\SS$. The  vector   
\ben
|h\rangle = \phi(0) | \Omega \rangle = \phi_{-h} | \Omega \rangle. 
\een
is called highest weight vector; we assume normalizations such that $\langle h | h \rangle = 1$. 
(ii) implies that $L_0|h\rangle=h|h\rangle$ and $L_n|h\rangle = 0$ for all $n>0$, and furthermore that $e^{iT(f)} \phi(z) e^{-iT(f)} = \rho'(z)^h \phi(\rho(z))$
in the sense of distributions in $z \in \SS$
where $\rho \in {\rm Diff}(\SS)$ is the diffeomorphism generated by the vector field ${\sf f}=f(z) \frac{\d}{\d z}$, see \eqref{rel1}.
Furthermore it is well-known (see e.g. \cite{carpi} for a mathematical account) that (i), (ii) and (iii) imply the operator product expansion
\ben
-2\pi \, T(w) \phi(z) |\Psi \rangle = \frac{h}{(w-z)^2} \phi(z) |\Psi \rangle + \frac{1}{w-z} \frac{\d}{\d z}\phi(z) |\Psi \rangle + \dots, \quad 
\text{if $e^{-s}< |z| < |w| < 1$}
\een
which is valid in the Hilbert space topology e.g. for vectors $|\Psi \rangle \in \D(e^{sL_0})$ and in the limit as $w \to z$. The dots represent terms of 
order $O(|w-z|^0)$. 

Now let $f \in C^\infty(\SS)$, real in the sense that $\Gamma f = f$, and compactly supported in a closed interval $I \subset \SS$. As in the previous section, define
\ben
G_h(z) = \langle e^{iT(f)} \rangle^{-1}_h 
\begin{cases}
\langle T(z) e^{iT(f)} \rangle_h & \text{if $|z|>1$,}\\
\langle e^{iT(f)} T(z) \rangle_h & \text{if $|z|<1$,}
\end{cases}
\een 
where $\langle A \rangle_h = \langle h | A h \rangle$. This function is again well defined and analytic on 
$\CC \setminus ({\rm supp}(f) \cup \{0\})$. Indeed, analyticity inside $\DD^+  \setminus \{0\}$ follows 
expanding out $T(z)$ in a power series in $z$, commuting the Virasoro generators $L_n$ through $\phi_{h}$ via (ii) and using the energy bound \eqref{poly} as 
well as $L_n |\Omega \rangle = 0$ for $n \ge -2$.  The 
operator product expansion of the quasi primary field implies that 
\ben
G_h(z) = -\frac{h}{2\pi} z^{-2} + O(z^{-1}) \quad \text{when $z \to 0$,}
\een
at the origin. The function is $G_h$ is also analytic on $\DD^-$ by a similar argument
and $G_h(z) = O(z^{-2})$ when $z \to \infty$ follows from the fact that $L_{n} |h\rangle = 0$ for $n \ge 1$. Again, we may argue using the edge-of-the-wedge
theorem that $G_h$ can be continued analytically across the circle $\SS$ where $f=0$. 

As above, we next define the two univalent holomorphic functions $w^\pm: \DD^\pm \to \Delta^\pm$ from the inside/outside of the circle onto the inside/outside $\Delta^\pm$ of a Jordan curve $C$, such that their respective boundary values on the circle $\SS$ satisfy the junction condition \eqref{jump2}, and which together define a corresponding univalent analytic function $w: \CC \setminus I \to \CC$.  Furthermore, similarly as above, we define the meromorphic function $H_h(w)$ on $\CC \setminus C$ as in \eqref{Hdef}, with boundary values on $C$ as 
in \eqref{Hdef1}. As before, $H_h(w)$ extends, in fact, to a meromorphic function on $\CC$ with a pole only at $w=0$, where it behaves as 
\ben
H_h(w) = -\frac{h}{2\pi} w^{-2} + O(w^{-1}) \quad \text{when $w \to 0$,}
\een
and as $H_h(w) = O(w^{-2})$ when $w \to \infty$. (These last statements follow from the behavior of $G_h$ at $z=0,\infty$ and
and the facts that $z=0$ gets mapped under $w^+$ to zero with unit differential at $0$, and that $z=\infty$ gets mapped under $w^-$ to infinity with unit differential on the real axis at infinity.) 
It follows that $w^2 H_h(w)$ is a bounded analytic function on $\CC$ and hence constant, thus 
$H_h(w) = -\tfrac{h}{2\pi}w^{-2}$. 

Reinstating the definition of $H_h$ in terms of $G_h$ and using the chain rule for the Schwarzian derivative, we obtain in a similar way as before
\ben
G^-_h(z) = -\frac{c}{24 \pi} Sw^-(z)  -\frac{h}{2\pi} \frac{1}{ w^-(z)^{2}} \left( \frac{\d w^-(z)}{\d z} \right)^2
\een
for $z \in \SS$, where $w^-$ is as in \eqref{wdef}. 
We can obviously repeat the same construction with the $\Gamma$-invariant test-function $tf(z), t \in \RR$, resulting in a $t$-dependent function $w(t,z)$ and 
a $t$-dependent function $G_h(t,z)$. As before, this leads to the formula
\ben\label{result3}
\langle e^{iT(f)} \rangle_h = \exp \left\{ \frac{1}{2\pi i} \int_0^1  \d t \int_{\SS} \d z \, f(z) \left( \frac{c}{12} Sw^-(t,z) +
h \left[\frac{\d}{\d z} \log w^-(t,z) \right]^2
\right) \right\} , 
\een
which is the analogue of the result \eqref{result} for highest weight states.

\subsection{Examples}

Here we illustrate our welding construction by giving an explicit derivation of the probability distributions $\d \nu_f$ of $T(f)$ (or $\Theta(f)$ for the CFT on the real line) in the vacuum state for the infinite family of test-functions $f_n$ on $\SS$ equal to 
\ben
f_n(z) = \frac{1}{2n}(z^{-n+1} - z^{n+1}) = \frac{1}{2n}(l_{-n}(z) - l_{n}(z)), 
\een
where $n \in \mathbb N$, giving the smeared fields
\begin{equation}
T(f_n) = \frac{1}{2in}(L_{-n}-L_n).
\end{equation}
The corresponding vector field ${\sf f}_n$ defined as in \eqref{ff} is real because $f_n(z) = \Gamma f_n(z) = -z^2 \overline{f_n(z)}$, and in fact 
\ben
{\sf f}_n = \frac{1}{2n}({\sf l}_{-n} - {\sf l}_{n}) = -\frac{1}{n} \sin(n \theta) \frac{\d}{\d \theta}.
\een
It is clear that this vector field has $2n$ zeros located at $z_k=e^{i\pi k/n}, k=0, 1, \dots, 2n-1$. Therefore, the corresponding 1-parameter group 
$\rho_t$ of diffeomorphisms of $\SS$ generated by ${\sf f}_n$ has the $2n$ fixed points $z_k$, too. Explicitly, we can write
\ben\label{rhot}
\rho_t(z) = e^{i\pi k/n} \left[(-1)^k \frac{z^n \cosh \t + \sinh \t}{z^n \sinh \t + \cosh \t} \right]^{\frac{1}{n}}
\quad \text{for $\pi k/n \le {\rm arg}(z) < \pi(k+1)/n$.}
\een
For $n=1$ this corresponds to a M\" obius transformation leaving $z=1, -1$ fixed. Since M\" obius transformations leave the vacuum state invariant, 
the probability distribution $\d \nu_n \equiv \d \nu_{f_n}$ of the stress tensor $T(f_n)$ is trivial if $n=1$, in the sense that  
\ben 
\d \nu_n(\lambda) = \delta(\lambda) \d \lambda \quad \text{for $n=1$}. 
\een 
This is easily understood because $T(f_1)=(2i)^{-1}(L_{-1}-L_{1})$ has $|\Omega\rangle$ as an eigenstate of zero eigenvalue. In the following, we therefore assume $n>1$. 

\begin{figure}
	\begin{center}\includegraphics[scale=0.7]{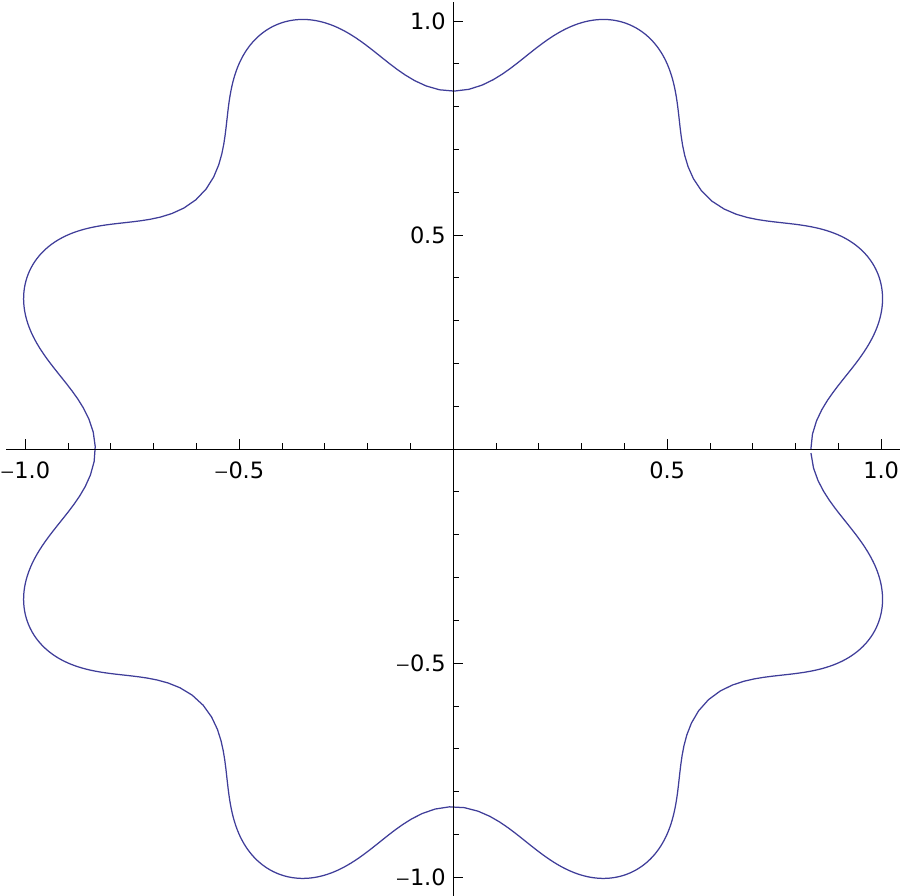}\end{center}
	\label{fig:plot}\caption{The Jordan curve $C$ separating $\Delta^+$ (inner region) and $\Delta^-$ (outer region) for $n=8$ and $t>0$. At $t=0$, $C$ is the unit circle, with bumps becoming more pronounced as $t$ increases.}
\end{figure}

Next, consider the conformal welding problem \eqref{jump2} for the diffeomorphism $\rho_t$
given by \eqref{rhot}. It is solved by the univalent holomorphic functions $w^+_n: \DD^+ \to \Delta^+$ and 
$w^-_n: \DD^- \to \Delta^-$ given by 
\ben 
w_n^+(z) = 
z \left(\cosh \t \right)^{-\frac{2}{n}} \left[ 1 + z^n \tanh \t \right]^{-\frac{1}{n}}
\een
and 
\ben 
w_n^-(z) = 
z  \left[ 1 - z^{-n} \tanh \t \right]^{\frac{1}{n}}
\een
for $z \in \DD^+$ and $z \in \DD^-$ respectively. Our way of writing $w^\pm_n$ shows in each case that it is holomorphic in $\DD^\pm$ as the 
term in square brackets is bounded away from the negative real axis in either case due to $|\tanh \t |<1$ and thus bounded away from the branch cut of the $n$-th root. The Jordan curve $C$ separating the domains $\Delta^\pm$ is drawn in fig.~\ref{fig:plot}.

In order to check that the above formulas give a solution to the welding problem \eqref{jump2} for $f_n$, one may argue as follows. First, we note that 
$\rho_t$ is invariant under a rotation by an angle of $2\pi/n$, or in other words $\rho_t(e^{i2\pi/n} z) = e^{i2\pi/n}  \rho_t(z)$. It then follows that $z \mapsto
e^{-i2\pi/n}w_n^\pm(e^{i2\pi/n} z)$ are solutions to the same welding problem mapping $z=0$ and the point at infinity on the real axis to themselves with unit derivative. 
By the uniqueness of the solution to the welding problem, we must therefore have $e^{-i2\pi/n}w_n^\pm(e^{i2\pi/n} z) = w^\pm_n(z)$, i.e. the solutions have 
to be periodic with period $2\pi/n$. Thus we can unambiguously define holomorphic univalent functions $v^+: \DD^+ \to \Delta^+$ and 
$v^-: \DD^- \to \Delta^-$ by 
\begin{equation}
v^\pm(z) = \begin{cases} [w^\pm(z^{1/n})]^n & 0\le\arg (z)<\pi \\ [w^\pm(e^{i\pi/n}(-z)^{1/n})]^n & 
\pi\le\arg(z)<2\pi,\end{cases}
\end{equation} 
and these have to be solutions of the welding problem 
$v^+ = v^- \circ g_t$ on $\SS$ where $g_t$ is the M\" obius transformation $g_t(z) = \frac{z \cosh \t + \sinh \t}{z \sinh \t + \cosh \t}$. Since 
M\" obius transformations extend to holomorphic univalent functions of $\DD^\pm$, the solution of this welding problem can be constructed trivially, with 
\begin{equation} 
v^-(z)=z-\tanh \t,\qquad v^+(z) = \frac{z\sech^2\t}{1+z\tanh\t}
\end{equation}
and then going back from $v^\pm$ to $w^\pm_n$ gives the above solution. 

By direct calculation, the Schwarzian derivative is given by 
\ben
Sw^-_n(z,t) = \frac{n^2-1}{2z^2} \left(
-1 + \frac{1}{(1- z^{-n} \tanh \t)^2}
\right) . 
\een
When $t \ge 0$, it has poles at $z=0, z=e^{i2\pi k/n}(\tanh \t)^{1/n}, k=0, 1, \dots, n-1$ inside the unit disk, and similarly for $t \le 0$. 
For definiteness, we now assume that $t \ge 0$. The other case is treated with a symmetry argument based on $\overline{\langle \exp itT(f_n)\rangle} = 
\langle \exp [-itT(f_n)] \rangle$ since $f_n$ is real.
Application of the residue theorem then gives 
\ben 
\frac{1}{2\pi i} \int_\SS f_n(z) Sw^-_n(z,t) \d z =  -\frac{n^2-1}{2n} \tanh \t . 
\een
We now substitute this into the formula \eqref{result} applied to the smearing function $f_n$. 
This shows that the Fourier transform of the probability distribution $\d \nu_n$ for $T(f_n)$ in the vacuum 
satisfies, for $n>1$, 
\ben
\int_\RR e^{it\lambda} \d \nu_n(\lambda) = \left( \sech \t \right)^p, 
\qquad p=\frac{c}{12} \left( n - \frac{1}{n} \right),
\een 
which is the characteristic function of a {\em generalized hyperbolic secant distribution}~\cite{Harkness:1968}. An inverse Fourier transform \cite[3.985]{GradshteynRyzhik:5thed} gives
\ben
\d \nu_n(\lambda) = 
\frac{2^{p-1}}{\pi\Gamma(p)}\left|\Gamma\left(\frac{p}{2}-i\lambda\right)\right|^2\d \lambda .
\een
The probability distribution function has a single peak centred at $\lambda=0$, about which it is symmetric; the breadth of the distribution increases with $n$.
The same probability distribution is obtained for $g_n(z) = \frac{i}{2n}(l_n(z)+l_{-n}(z))$ which is equivalent to the vector field ${\sf g_n} = \frac{1}{n}
\cos (n\theta) \frac{\d}{\d \theta}$, since ${\rm rot}_{\pi/2}^* {\sf g_n} = {\sf f_n}$, so $e^{itT(f_n)} = e^{-i\pi L_0/2} e^{itT(g_n)} e^{i\pi L_0/2}$ (by \eqref{rel1}) and since the vacuum vector is invariant under $e^{i\pi L_0/2}$.  

As a consistency check on these results, let us note that the second moment of $T(f_n)$ in the vacuum state may be computed directly, using the Virasoro relations and $L_m\Omega=0$ for $m\ge 0$. One finds
\begin{equation}
\ip{\Omega}{T(f_n)^2\Omega} =\frac{1}{4n^2}\ip{\Omega}{[L_n,L_{-n}] \Omega} = 
 c \frac{n^2-1}{48n}.
\end{equation}
On the other hand, the second moment is minus the second derivative of the characteristic function at $t=0$, and on noting
\begin{equation}
\left(\sech\t\right)^p = \left(1 - \frac{t^2}{8} + O(t^4)\right)^p = 1 - \frac{ct^2(n^2-1)}{96n} + O(t^4),
\end{equation}
we have agreement with our direct calculation. 

Finally, we may also calculate the probability distribution $\d\nu_{n,h}(\lambda)$ for $T(f_n)$ in a highest weight state $|h \rangle$ rather than the vacuum, now using \eqref{result3} in place of \eqref{result}. This gives again a generalized hyperbolic secant distribution for the probability measure $\d\nu_{n,h}(\lambda)$, where the parameter $p$ is now
\ben
p=\frac{c}{12} \left( n - \frac{1}{n} \right) + \frac{2h}{n}.
\een
Again, this correctly yields the second moment of $T(f_n)$ in this state, providing a consistency check.
 
 \section{Part 2: Probability distributions via moment generating functions}\label{sec:MGF}
\subsection{General theory}

We briefly recall how the probability distribution may be recovered from its moments, working with the stress energy density on the line. Let $f$ be any real-valued test function, so that $\Theta(f)$ is a self-adjoint operator with an associated projection-valued measure $P_f(\d \lambda)$ on the real line. Recall that the vacuum probability distribution of $\Theta(f)$ corresponds to the measure 
\begin{equation}
\d \nu_f(\lambda) = \ip{\Omega}{P_f(\d \lambda)\Omega}
\end{equation}
and of course the distribution in any other state is obtained by setting it in place of $\Omega$. The $n$'th moment of $\nu_f$ is 
\begin{equation}
m_n = \int_\RR \lambda^n\, \d\nu_f(\lambda) 
\end{equation}
and by functional calculus, one finds
\begin{equation}
m_n = \int_\RR \lambda^n \ip{\Omega}{P_f(\d \lambda)\Omega}=\ip{\Omega}{\Theta(f)^n\Omega}
\end{equation}
as is well-known. 

In~\cite{FewsterFordRoman:2010} it was shown that the moment generating function
\begin{equation}
M[\mu f]=\sum_{n=0}^\infty \frac{\mu^n}{n!}\ip{\Omega}{\Theta(f)^n\Omega}
\end{equation}
can be expressed as
\begin{equation}\label{eq:MfromW}
M[\mu f]= \exp W[\mu f],
\end{equation}
where
\begin{equation}\label{eq:Wfromf}
W[\mu f]= \int_0^\mu \d\lambda\, (\mu-\lambda) \ip{\Omega}{\Theta(f_\lambda)^2\Omega} 
\end{equation}
and $f_\lambda$ solves the flow equation
\begin{equation}\label{eq:flow}
\frac{\d f_\lambda}{\d\lambda} = f_\lambda\star f_\lambda,\qquad f_0=f,
\end{equation}
with the $\star$ operation defined by
\begin{equation}
\label{fstf}
(f\star f)(u):=\int_{-\infty}^\infty \d w \frac{f(w)f'(u)-f'(w)f(u)}{2\pi(w-u)}.
\end{equation}
This result was reached as a consequence of the CFT Ward identities, which yield recursion relations for the moments and related quantities. In the first instance~\eqref{eq:MfromW} is to be understood as an equality of formal power series in $\mu$; however, 
if $W[\mu f]$ is holomorphic in a neighbourhood of the origin, it becomes an identity of functions. As we also have, formally,
\begin{equation}
M[\mu f]= \int_\RR e^{\mu\lambda} \, \d\nu_f(\lambda)  ,
\end{equation}
the process of recovering the probability distribution becomes one of inverting a 
Laplace transform. This outline must be supplemented by conditions to guarantee
that the moments uniquely determine the distribution, for example, the Hamburger
condition that $|m_n|\le C D^n n!$ for constants $C$ and $D$. If $f$ is nonnegative, the existence of a finite QEI bound \eqref{QEI} (see~\cite{Fe&Ho05} for details) implies that the vacuum distribution is supported on a half-line and therefore the Stieltjes condition $|m_n|\le C D^n (2n)!$ would also suffice to guarantee uniqueness of the reconstructed distribution.  See~\cite{Simon:1998} for an exposition of these and other facts concerning moment problems in general.

A specific solution to~\eqref{eq:flow} was given in~\cite{FewsterFordRoman:2010} for the case where $f$ is a Gaussian, namely
\begin{equation}
f_\lambda(u) = \frac{\tau\sqrt{\pi}}{\pi\tau^2-\lambda}e^{-(u/\tau)^2}.
\end{equation}
Using the expression
\begin{equation}
\ip{\Omega}{\Theta(f_\lambda)^2\Omega} = \frac{c}{48\pi^2}\int_0^\infty \d\omega\, \omega^3 |\hat{f_\lambda}(\omega)|^2,
\end{equation}
the moment generating function was obtained in closed form as 
\begin{equation}
M[\mu f]= \left(\frac{e^{-\mu/(\pi\tau^2)}}{1-\mu/(\pi \tau^2)}\right)^{c/24}.
\end{equation}
This can be compared with the moment generating function
\begin{equation}
M_{\alpha,\beta,\sigma}(\mu) = e^{-\mu\sigma}(1-\mu\beta)^{-\alpha}
\end{equation}
of the shifted Gamma distribution~\eqref{eq:shiftedGammapdf}, from which it follows
that $\mu\mapsto M[\mu f]$ is the shifted gamma distribution $\d\nu_f(\lambda)$ given by~\eqref{eq:shiftedGammapdf} 
with parameters~\eqref{eq:Gaussian_pars}. Given that the moments satisfy the Hamburger moment criterion, this is the unique probability distribution with this moment generating function. As already mentioned, $-\sigma$ is the QEI lower bound~\cite{Fe&Ho05} for Gaussian averaging. 

No other solutions to the flow equation were known when~\cite{FewsterFordRoman:2010} was written so the Gaussian result seemed perhaps to be an isolated curiosity. We will describe two new infinite families of solutions below in sec. \ref{sec:lorentz} and \ref{sec:gamma},
and then explain in sec.~\ref{sec:Gibbs} how the above approach can be extended to a Gibbs state on a CFT in the circle picture.

\subsection{Lorentzian family}
\label{sec:lorentz}
For $b>0$, consider functions of the form
\begin{equation}
f(u)= \frac{a}{(b^2+u^2)^n},
\end{equation}
with derivative
\begin{equation}
f'(u)= -\frac{2nu a}{(b^2+u^2)^{n+1}}.
\end{equation}
For simplicity, we take $n\ge 1$ to be integer, but large parts of the analysis go through without change for arbitrary real $n>0$. Noting that 
\begin{equation}
\frac{f(w)f'(u)-f'(w)f(u)}{w-u} = \frac{2nb^2a^2}{(b^2+w^2)^{n+1}(b^2+u^2)^{n+1}} - \frac{2nuwa^2}{(b^2+w^2)^{n+1}(b^2+u^2)^{n+1}},
\end{equation}
of which the second term is evidently odd in $w$, 
we find
\begin{equation}
(f\star f)(u) = \frac{n a^2\kappa_{n+1}}{(b^2+u^2)^{n+1}b^{2n-1}},
\end{equation}
where
\begin{equation}
\kappa_n = \int_{-\infty}^\infty \frac{\d x}{\pi} \frac{1}{(1+x^2)^{n}} = \frac{\Gamma(n-1/2)}{\Gamma(n)\sqrt{\pi}} .
\end{equation} 
Equivalently, one has
\begin{equation}
\kappa_1=1,\qquad  \kappa_n=\frac{8}{4^n}\binom{2n-3}{n-1},~\quad (n\ge 2).
\end{equation}
Now set $f_\lambda$ equal to $f$ with $b$ replaced by $b_\lambda$. Noting that
\begin{equation}
\frac{\d f_\lambda}{\d\lambda} =-\frac{2n b_\lambda a}{(b_\lambda^2+u^2)^{n+1}} \frac{\d b_\lambda}{\d\lambda},
\end{equation}
this \emph{ansatz} solves the flow equation provided that
\begin{equation}
b_\lambda^{2n}\frac{\d b_\lambda}{\d\lambda} = -\frac{\kappa_{n+1} a}{2},
\end{equation}
which has the unique solution
\begin{equation}
b_\lambda = b_0\left(1-\frac{(2n+1)\kappa_{n+1} a\lambda}{2b_0^{2n+1}}\right)^{1/(2n+1)}.
\end{equation}
If $f_0$ is normalized to have unit integral, then
\begin{equation}
a   = \frac{b_0^{2n-1}}{\pi\kappa_n} 
\end{equation}
so one has
\begin{equation}
b_\lambda = b_0\left(1-\frac{(4n^2-1)\lambda}{4n \pi b_0^{2}}\right)^{1/(2n+1)}.
\end{equation}

Now 
\begin{equation}
\langle \Theta(f)^2\rangle = \frac{ca^2}{48\pi^2 b^{4n+2}}K_n,
\end{equation}
where 
\begin{equation}
K_n = \int_0^\infty y^3\, \left|\mathcal{F}[x\mapsto (1+x^2)^{-n}](y)\right|^2\,\d y\\
\end{equation}
is evaluated in the Appendix as
\begin{equation}
K_n=\frac{2\pi^2 n^2}{16^n}\binom{2n-1}{n}\binom{2n+1}{n}.
\end{equation}
Inserting the normalization for $a$ and the solution to the flow equation yields
\begin{equation}
\langle \Theta(f_\lambda)^2\rangle = \frac{cK_n}{48\pi^4 \kappa_n^2 b_0^{4}}A_n(\lambda)^2,
\end{equation}
where
\begin{equation}\label{eq:betanLor}
A_n(\lambda) = \frac{\beta_n}{\beta_n-\lambda},\qquad \beta_n = \frac{4n\pi b_0^2}{4n^2-1}.
\end{equation}
It follows that
\begin{equation}
W[\mu f] = \frac{cK_n}{48\pi^4 b_0^{4}} \left( \beta_n^2\log A_n(\mu) - \beta_n\mu\right) = 
\frac{cK_n\beta_n^2}{48\pi^4 b_0^4}  \log \frac{e^{-\mu/\beta_n}}{1-\mu/\beta_n}
\end{equation} 
and hence the moment generating function is again that of a shifted gamma distribution \eqref{eq:shiftedGammapdf}
with parameters $\alpha_n$, $\beta_n$ and $\sigma_n$, where $\beta_n$ was given above and
\begin{equation}\label{eq:alphansigmanLor}
\alpha_n = \frac{cK_n\beta_n^2}{48\pi^4\kappa_n^2 b_0^{4}}, \qquad \sigma_n = \frac{\alpha_n}{\beta_n}.
\end{equation}
Substituting the explicit formulae for $K_n$ and $\kappa_n$ and simplifying, one finds
\begin{equation}
\alpha_n = \frac{c n^2}{12(2n+1)(n+1)}, \qquad \sigma_n=\frac{c n(2n-1)}{48(n+1)\pi b_0^2}.
\end{equation}

The parameter $\sigma_n$ can be compared with the QEI bound \eqref{QEI}.  To this end, we compute
\begin{equation}
\left(\frac{\d}{\d u}\sqrt{f(u)}\right)^2 = \frac{an^2 u^2}{(b^2+u^2)^{n+2}} = 
\frac{an^2}{(b^2+u^2)^{n+1}} - \frac{ab^2 n^2}{(b^2+u^2)^{n+2}}
\end{equation}
so
\begin{align}
\frac{c}{12\pi}\int_{-\infty}^\infty\left(\frac{\d}{\d u}\sqrt{f(u)}\right)^2\,\d u &= \frac{an^2 c}{12  b^{2n+1}} \left(  \kappa_{n+1} -  \kappa_{n+2}\right) = 
\frac{n^2 c\left(  \kappa_{n+1} -  \kappa_{n+2}\right)}{12\pi\kappa_n b_0^2}\nonumber\\
&= \frac{c n(2n-1)}{48(n+1)\pi b_0^2} = \sigma_n
\end{align}
on inserting the normalization for $f=f_0$. As is required on general grounds, 
the QEI bound therefore coincides with the infimum of the support of the vacuum probability distribution.

\subsection{Inverse Gamma sampling}
\label{sec:gamma}

For $\mu\in\RR$, $b>0$, $\gamma>1$, let
\begin{equation}\label{eq:invgam}
f(u) = \vartheta(u-\mu)a \frac{e^{-b/(u-\mu)}}{(u-\mu)^{\gamma}}.
\end{equation}
which is normalized to unit integral if one has 
\begin{equation}\label{eq:invgamnorm}
a=\frac{b^{\gamma-1}}{\Gamma(\gamma-1)}.
\end{equation} 
With this normalization, $f$ is the probability density function of an inverse gamma distribution, supported on $[\mu,\infty)$.

As before, the first step is to calculate $f\star f$. Starting from
\begin{equation}\label{eq:invgamder}
f'(u) = \frac{f(u)}{(u-\mu)^2}\left( b- \gamma(u-\mu)\right),
\end{equation}
a calculation gives
\begin{equation}
\frac{f(w)f'(u)-f'(w)f(u)}{w-u} =
 \frac{f(u)}{(u-\mu)^2} \frac{b f(w)}{w-\mu} + \frac{f(u)}{u-\mu} f'(w)
\end{equation}
so 
\begin{equation}
(f\star f)(u) = \frac{f(u)}{(u-\mu)^2} \int_\mu^\infty \frac{b f(w)}{w-\mu}\,\frac{\d w}{2\pi},
\end{equation}
and the integral is easily evaluated, giving
\begin{equation}
(f\star f)(u) = \frac{a \Gamma(\gamma) f(u)}{2\pi b^{\gamma-1}(u-\mu)^2}. 
\end{equation}

On the other hand, 
\begin{equation}
\frac{\d f(u)}{\d\mu} = -\frac{f(u)}{(u-\mu)^2}\left(b+(u-\mu)\frac{\d b}{\d\mu} -\gamma (u-\mu)\right) =
-\frac{b f(u)}{(u-\mu)^2},
\end{equation}
provided we take $b=\gamma (\mu+\delta)$ with $\delta$ a constant. Note that we must stay in the regime where $\mu+\delta>0$. 

Now make the \emph{ansatz} $f_\lambda=f$ with $\mu=\mu(\lambda)$, $b=\gamma (\mu+\delta)$, but keeping $a$ constant. Then the flow equation is satisfied if
\begin{equation}
\frac{\d\mu}{\d\lambda} = -\frac{a\Gamma(\gamma)}{2\pi\gamma^\gamma (\mu+\delta)^\gamma}
\end{equation}
which has solution
\begin{equation}
\mu(\lambda) +\delta = (\mu_0+\delta) \left(1 - \frac{a\Gamma(\gamma+2)\lambda}{2\pi[\gamma(\mu_0+\delta)]^{\gamma+1}}\right)^{1/(\gamma+1)}
\end{equation}
Evidently we may absorb $\delta$ into $\mu$ and $\mu_0$ and we therefore do so.

To evaluate $\langle\Theta(f_\lambda)^2\rangle$ we need the Fourier transform $\hat{f}$. 
As only the modulus enters, we may set $\mu=0$ and work with $\vartheta(u)e^{-b/u}/u^\gamma$.
Then
\begin{equation}
\langle\Theta(f)^2\rangle = \frac{c a^2 K_\gamma}{48\pi^2 b^{2\gamma+2}} 
\end{equation}
where
\begin{align}
K_\gamma &:= \int_0^\infty y^3\, \left|\mathcal{F}[x\mapsto \vartheta(x) x^{-\gamma} e^{-1/x}](y)\right|^2\,\d y \\
&= \frac{1}{2}\Gamma(\gamma+1)\Gamma(\gamma+3).
\end{align}
The evaluation of this integral is described in the Appendix.

Inserting the normalization for $a$ and the solution to the flow equation,
\begin{equation}
\langle \Theta(f_\lambda)^2\rangle = \frac{cK_\gamma}{48\pi^2 \Gamma(\gamma-1)^2 b_0^{4}}A(\lambda)^2,
\end{equation}
where
\begin{equation}
A(\lambda) = \frac{\beta}{\beta-\lambda},\qquad \beta = \frac{2\pi b_0^2}{(\gamma^2-1)\gamma}.
\end{equation}
This leads to a shifted gamma distribution \eqref{eq:shiftedGammapdf} with parameters $\alpha$, $\beta$ and $\sigma$,
where $\beta$ is as above and
\begin{equation}
\alpha = \frac{c\beta^2 K_\gamma}{48\pi^2 \Gamma(\gamma-1)^2 b_0^{4}} = 
\frac{c K_\gamma}{12 \Gamma(\gamma+2)^2 }, \qquad \sigma = \frac{\alpha}{\beta} = \frac{cK_\gamma}{24\pi\Gamma(\gamma+2)\Gamma(\gamma-1)b_0^2}.
\end{equation}
Substituting the value of the constant $K_\gamma$, we have, overall:
\begin{equation}
\alpha = \frac{c(\gamma+2)}{24(\gamma+1)}, \qquad 
\beta = \frac{2\pi b_0^2}{(\gamma^2-1)\gamma}, \qquad 
\sigma = \frac{c(\gamma+2)(\gamma-1)\gamma}{48\pi b_0^2}.
\end{equation}
The value of $\sigma$ agrees exactly with the QEI bound calculated for~\eqref{eq:invgam} with normalization~\eqref{eq:invgamnorm},
which is easily calculated using $(d\sqrt{f(u)}/du)^2=f'(u)^2/(4f(u))$ 
together with~\eqref{eq:invgamder} and the same integrals used to compute the normalization.

We have discussed two distinct infinite families of test functions on the light ray for which the vacuum probability distribution can be obtained in closed form. These can immediately be translated into examples on the circle, using the correspondence~\eqref{real}.
Furthermore, the $\rm{PSL}(2,\RR)$-covariance of the stress-energy tensor immediately provides further new examples. To be specific, if $\rho\in\rm{PSL}(2,\RR)$, the Schwarzian derivative in~\eqref{rel1} vanishes and one finds that 
$T(f)$ is unitarily equivalent to $T(f_\rho)$ where $f_\rho(z)=f(\rho^{-1}(z))\rho'(\rho^{-1}(z))$. As the unitary transformation leaves the vacuum vector invariant, the vacuum probability distributions of $T(f)$ and $T(f_\rho)$ are identical.
This also applies on the line: $\Theta(f)$ and $\Theta(f_\rho)$ have the same distribution if $\rho$ is an orientation-preserving M\"obius transformation mapping the real line to itself,
writing $f_\rho(u)=f(\rho^{-1}(u)) \rho'(\rho^{-1}(u))$. 

As a particular example, we note that $\rho(u)= -1/u$ converts the inverse Gamma test function 
given by \eqref{eq:invgam} with $\mu=0$ to the Gamma density function
$f_\rho(u)=\vartheta(-u)a (-u)^{2+\gamma} e^{bu}$ on the negative half-line, and one may equally consider shifted or reflected versions thereof. Thus these test functions (which are smooth except at $u=0$, where they have finite order of differentiability) also have shifted Gamma vacuum probability distributions.

By mapping the positive real line to a bounded interval one obtains an example of a compactly supported test function on $\RR$ (smooth except at one endpoint, where there is only finite differentiability) with a shifted Gamma distribution. Now consider two such test functions $f_1$ and $f_2$ with disjoint support. As $\Theta(f_1)$ and $\Theta(f_2)$ commute, 
the vacuum distribution for $\Theta(f_1+f_2)$ is simply the convolution of the two individual 
vacuum distributions. This will again be a shifted Gamma if the QEI bounds of $f_1$ and $f_2$ are equal. The distribution for other (not necessarily positive) linear combinations of the $f_i$ can be determined as well.

 \subsection{Thermal states}\label{sec:Gibbs}
 
The moment generating technique can be extended, in principle, to thermal states and presumably also to other special states. 
 As an illustration, we now look at the probability distribution of the smeared stress tensor $T(f)$ 
 on the circle $\SS$ in a Gibbs state, 
 \ben
 \langle A \rangle_\beta = Z(\beta)^{-1} {\rm Tr}\, (A e^{-\beta L_0}), \quad \beta > 0, \quad Z(\beta) = {\rm Tr}\, e^{-\beta L_0} ,  
 \een
 where we assume that $Z(\beta) < \infty$.
 The idea is as in \cite{FewsterFordRoman:2010} to use the conformal Ward identities (now of the Gibbs state) in order to 
 get recursive relations between the moments, now given by $m_n = \langle T(f)^n \rangle_\beta$, where $f$ is a real (in the sense 
 $\Gamma f = f$) test-function on $\SS$. To simplify the formulas, we are going to use the coordinate 
 $z=e^{i\theta}$ on $\SS$, where $\theta \in (0,2\pi]$ is periodic, and we define, by abuse of notation
 \ben
 T(\theta) \equiv -2\pi z^2T(z), \quad z=e^{i\theta}, 
 \een
 which corresponds to the transformation law of the stress tensor (up to the central term and a factor of $2\pi$). 
 If we simultaneously change $f(z)$ to $f(\theta) \equiv -ie^{-i \theta} f(e^{i\theta})/2\pi$ (up to $2\pi$ the transformation law of the vector field \eqref{ff} on $\SS$), then 
 we can write $T(f) = \int_0^{2\pi} T(\theta) f(\theta) \d \theta$ and $f(\theta)$ is now real-valued in the ordinary sense. 
 
 With these conventions understood for the remainder of this section, the conformal Ward identities are most conveniently expressed in terms of 
 the distributions 
 \ben
 G_{n,\beta}(\th_1, \dots, \th_n) = 
 {\rm Tr}\left( T(\th_1) \cdots T(\th_n) e^{-\beta L_0 } \right)
 \een
 were each $\th_i \in (0,2\pi]$ is the periodic coordinate on $\SS$ just defined.
 Evidently, $G_{0,\beta}=Z(\beta)$, while one finds $G_{1,\beta}(\theta)\equiv -\d Z/\d\beta$ (independent of $\theta$) as a consequence of cyclicity
 of the trace and the identity $L_k e^{-\beta L_0}=e^{-\beta k}e^{-\beta L_0}L_k$. 
 Adapting \cite{felder} to our conventions, the Ward identities read, for $n\ge 1$:
 \ben
 \begin{split}
 & G_{n+1,\beta}(\th,\th_1, \dots, \th_n) \\
 =&-\frac{c}{24 \pi} \sum_{j=1}^n \varphi'''(\th_j-\th + i0)   G_{n-1,\beta}(\th_1, \dots, \hat \th_j, \dots, \th_n) -\frac{\partial}{\partial \beta} G_{n,\beta}(\th_1, \dots, \th_n) \\
  &+ \frac{1}{2\pi}\sum_{j=1}^n \left[ 
 2\varphi'(\th_j-\th+i0) +  \varphi(\th_j-\th+i0) \frac{\partial}{\partial \th_j}
 \right] G_{n,\beta}(\th_1, \dots, \th_n).
 \end{split}
 \een
 Here, a caret on $\hat \th_j$ means omission and
 \ben
 \varphi(\th) = \sum_{n \in \ZZ, n\neq 0} \frac{e^{in\th}}{1-e^{-\beta n}}  = i \zeta(\th) + \frac{\eta_1(\beta)}{i\pi} \th  -\frac12  
 \een
 where $\zeta$ is the Weierstrass zeta function with full periods 
 $(2\pi,i\beta)$ on the real/imaginary axis, $\eta_1(\beta) = \zeta(\pi)$. Explicitly, in terms of Jacobi $\vartheta$-functions,
 \ben
\zeta(\th) = \frac{\vartheta_1'(\th | i\beta )}{\vartheta_1(\th |  i\beta)} + \frac{\eta_1(\beta)}{\pi} \th. 
 \een
 If we multiply the Ward identity with a product of $n+1$ factors of the test-function $f$ and integrate against 
 $\th,\th_j$, we obtain after a few trivial manipulations
 \ben\label{rec}
 \begin{split}
 G_{n+1,\beta}(f, \dots, f) =& -\frac{nc}{24 \pi} \left( \int_0^{2\pi} \int_0^{2\pi} \varphi'''(\th_1-\th_2+i0) f(\th_1) f(\th_2) \d \th_1 \d \th_2 \right)   G_{n-1,\beta}(f, \dots, f) \\
  &-\left( \int_0^{2\pi} f(\th) \d \th \right) \frac{\partial}{\partial \beta} G_{n,\beta}(f, \dots, f)
  + \sum_{j=1}^n  G_{n,\beta}(f, \dots, f \star_\beta f, \dots, f), 
 \end{split}
 \een
 where, analogous to \eqref{fstf},  
 \begin{equation}
\label{fstf1}
(f\star_\beta f)(\th):=\int_{0}^{2\pi} \frac{\d \th'}{2\pi}  [f(\th')f'(\th)-f'(\th')f(\th) ] \varphi(\th'-\th).
\end{equation}
Now we multiply \eqref{rec} by $\mu^n/n!$, we sum over $n$ from $1$ to infinity and we divide by the partition function $Z(\beta)$. We then get a partial differential 
equation for the moment generating function
\ben
M_\beta[\mu f] \equiv \sum_{n=0}^\infty \frac{\mu^n}{n!} \langle T(f)^n \rangle_\beta. 
\een
To solve this equation 
we consider a 1-parameter family $(f_\lambda, \beta_\lambda)$ of a periodic test function 
$f_\lambda$ on the interval $(0,2\pi)$ and a $\lambda$-dependent temperature parameter $\beta_\lambda$
solving the coupled flow equations
\ben
\frac{\d f_\lambda}{\d \lambda} = f_\lambda \star_{\beta_\lambda} f_\lambda, \quad
\frac{\d \beta_\lambda}{\d \lambda} = -\int_0^{2\pi} f_\lambda \d \th, 
\een
with the initial conditions $f_0 = f, \beta_0=\beta$. Then the partial differential equation for the moment generating function
becomes
\ben
\left( \frac{\partial}{\partial \mu} - \frac{\partial}{\partial \lambda} \right) \log M_{\beta_\lambda}[\mu f_\lambda] - 
\frac{\partial}{\partial \lambda} \log Z(\beta_\lambda) \\
= \mu \Phi_\lambda[f_\lambda], 
\een
where
\begin{equation}
\Phi_\lambda[f]:=-\frac{c}{24 \pi}  \int \varphi'''_\lambda(\th_1-\th_2+i0) f (\th_1) f (\th_2) \, \d \theta_1 \d \theta_2
\end{equation}
and the subscript on $\varphi_\lambda$ means that we should insert the $\lambda$-dependent temperature parameter $\beta_\lambda$, and where
the variables $\theta_i$ are integrated from $0$ to $2\pi$. The solution is with our boundary conditions,
\begin{equation}
\log M_{\beta_\lambda}[\mu f_\lambda] =\log (Z(\beta_{\mu+\lambda})/Z(\beta_\lambda)) + \int_\lambda^{\lambda+\mu} (\mu+\lambda-\lambda')\Phi_{\lambda'}[f_{\lambda'}]\, \d\lambda'
\end{equation}
 and after setting $\lambda=0$,
\ben
M_\beta[\mu f] = \frac{Z(\beta_\mu)}{Z(\beta)} \exp \left[
 \int_0^\mu (\mu-\lambda) \Phi_\lambda[f_\lambda]\,\d \lambda
\right].
\een
As in the vacuum case, the above formulas are, a priori, understood as formal series in $\mu$, the convergence of which will in general depend on our choice of $f$. Note that the exponent in this result is directly proportional to the central charge $c$. 

Our result can be put into a simpler form on noting that (suppressing the subscript $\lambda$) the Ward identity~\eqref{rec} with $n=1$ gives
\begin{align}
\Phi[f]&= \langle T(f)^2\rangle_\beta - \left(\int_0^{2\pi} f(\th)\, \d \th\right)^2
\frac{1}{Z}\frac{\d^2 Z}{\d\beta^2} + \int_0^{2\pi} (f\star_\beta f)(\th)\, \d \th
\frac{1}{Z}\frac{\d Z}{\d\beta} \notag\\
&= \langle T(f)^2\rangle_\beta^{C} - \left(\int_0^{2\pi} f(\th)\, \d \th\right)^2
\frac{\d^2}{\d\beta^2}\log Z + \int_0^{2\pi} (f\star_\beta f)(\th)\, \d \th
\frac{\d}{\d\beta}\log Z ,
\end{align}
where the connected two-point function is 
\begin{equation}
\langle T(f)^2\rangle_\beta^{C}=\langle T(f)^2\rangle_\beta-\langle T(f)\rangle_\beta^2.
\end{equation}
Along the flow, therefore, we have
\begin{equation}
\Phi_\lambda[f_\lambda] = \langle T(f_\lambda)^2\rangle_{\beta_\lambda}^{C} -
\frac{\d^2}{\d\lambda^2}\log Z(\beta_\lambda),
\end{equation}
and on noting that 
\begin{align}
\int_0^\mu (\mu-\lambda) \frac{\d^2}{\d\lambda^2}\log Z(\beta_\lambda)\d \lambda &= 
\log \frac{Z(\beta_\mu)}{Z(\beta)} - \mu \left.\frac{\d}{\d\lambda}\log Z(\beta_\lambda)\right|_{\lambda=0}\notag\\
&= \log \frac{Z(\beta_\mu)}{Z(\beta)} - \mu\langle T(f)\rangle_\beta,
\end{align}
we obtain
\begin{equation}
M_\beta[\mu f] =  \exp \left[\mu\langle T(f)\rangle_\beta + 
\int_0^\mu (\mu-\lambda) \langle T(f_\lambda)^2\rangle_{\beta_\lambda}^{C} \d \lambda \right],
\end{equation}
which generalises equations~\eqref{eq:MfromW} and~\eqref{eq:Wfromf} to the Gibbs state (for the CFT on the circle). 

Thus, the final answer for the moment generating function has a similar structure as in the vacuum case. Instead of one flow equation, we now 
need to solve a coupled pair of flow equations, and the kernel in $\star_\beta$ is more complicated than the corresponding kernel in the vacuum case, being an 
elliptic function rather than a rational function. As a consequence, it is presumably harder to find examples of functions $f$ where an explicit solution is available. As  in the vacuum case, we need the conditions of the Hamburger/Stieltjes moment problems to be satisfied in order 
to get a unique probability distribution of $T(f)$ from the moments in our Gibbs state.

 \section{Conclusions}\label{sec:conclusion}
 
The ability to find closed form expressions for the probability distribution of the smeared stress energy operator, at least in principle, makes 2-dimensional CFTs particularly important in this context. The novel examples of probability distributions for non-negative smearing functions $f$ are all  given by a shifted gamma-distribution, the parameters of which depended on the smearing function. Furthermore, in all cases the probability distributions were uniquely characterized by their moments in view of the Stieltjes/Hamburger theorems. 

Unfortunately, all our methods relied, in one way or another, on the powerful constraints imposed by conformal invariance in two dimensions. In fact, 
in the moment generating technique, one uses the conformal Ward identities, whereas in the welding technique, one uses the connection between Virasoro symmetry and diffeomorphisms of the real line (or circle). Thus, it seems that our methods have no straightforward generalization to theories without conformal invariance, or to higher spacetime dimensions.  

In fact, even for free massless fields in four dimensional spacetime, a closed form has appeared to be out of reach, and previous results on this theory have focussed on 
asymptotic behaviour of the moments of the distribution \cite{FewForRom:2012,FewFord:2015} or been obtained by numerically diagonalising the energy density operator~\cite{SchiaFewsFord:2018}. Among other things, 
it has been shown that the moments of the time-averaged energy density operator can grow very rapidly -- so rapidly, indeed, that they fall outside the scope of the Stieltjes and Hamburger theorems that guarantee the unique reconstruction of a distribution from its moments~\cite{Simon:1998}. One may nonetheless infer qualitative information about the tail of the probability distribution and this has confirmed independently by the numerical study~\cite{SchiaFewsFord:2018}. 
One may ask similar questions of other observables. In particular, a combination of analytical and numerical methods were used to examine the moments of the Wick square, averaged against the Lorentzian function, i.e., the $n=1$ member of our `Lorentzian family'~\cite{FewForRom:2012}. Using some novel combinatorics~\cite{FewSiem:2014} it was possible to compute the first 65 moments exactly, obtaining an exact match with the moments of a shifted Gamma distribution with parameters 
$\alpha=1/72$, $\beta=4\pi^2 b_0^2/3$, $\sigma =1/(96\pi^2 b_0^2)$.\footnote{Our parameter $b_0$ corresponds to $\tau$ in~\cite{FewForRom:2012}; note that Eq.~(10) of that reference reports the parameters for the probability distribution of a non-dimensionalised quantity that is $(4\pi\tau)^2$ times the averaged Wick square.} As is seen from~\eqref{eq:betanLor} and~\eqref{eq:alphansigmanLor}, these are the parameters appropriate to the $c=1$ CFT stress-energy probability distribution averaged against the Lorentzian and divided by $\pi$. An explanation for this can be given as follows: restricted to a timelike line, the massless field in $3+1$ dimensions can be expressed as an infinite tensor product of theories that are related to the $c=1$ free current~\cite[Sec.~8]{BuchholzDAntoniLongo}, under which correspondence the Wick square is mapped to the chiral stress tensor (divided by $\pi$). It is hoped to investigate further this intriguing direction elsewhere.

Finally, we should remark that quantities similar to those we have studied also appear in the problem of determining the so-called full counting statistics of heat flows in certain non-equilibrium states in CFT; for specific calculations in Luttinger model see e.g.~\cite{GawedTauber} and references therein. After this paper was posted, we have learned that welding methods also have, independently, been applied in that context in a forthcoming work by Gawedzki~\cite{Gawedzki_talk}.
 
\medskip
\noindent\emph{Acknowledgement} CJF thanks the Institute for Theoretical Physics at the University of Leipzig for kind hospitality during a visit when this work was commenced. We both thank the anonymous referee for useful suggestions and in particular for bringing Refs.~\cite{Baumann} and~\cite{BuchholzDAntoniLongo} to our attention.
	We also thank K Gawedzki for useful discussions during the BIRS meeting `Physics and Mathematics of Quantum Field Theory', July-August 2018, Banff.

\appendix
\section{Evaluation of $K_n$ and $K_\gamma$}

We evaluate two families of integrals needed in the text. First, consider
\begin{equation}
K_n :=\int_0^\infty \omega^3 |\widehat{g_n}(\omega)|^2\,\d \omega ,
\end{equation}
where $g_n(x)=(1+x^2)^{-n}$. Evaluation of the transform by residue methods gives, for $\omega>0$,
\begin{equation}
\widehat{g_{n+1}}(\omega) = \frac{\pi e^{-\omega}}{2^{2n}}\sum_{r=0}^n \binom{2n-r}{n}\frac{(2\omega)^r}{r!},
\end{equation}
so $K_{n+1}$ becomes an integral of a double sum, of which the integral
can be performed first to give
\begin{equation}
K_{n+1}=\frac{6\pi^2}{16^{n+1}}\sum_{r=0}^n\sum_{s=0}^n \binom{2n-r}{n}\binom{2n-s}{n}\binom{r+s+3}{r+3}\binom{r+3}{r}.
\end{equation}
The sum on $s$ may be performed using the identity
\begin{equation}
\sum_{s=0}^n \binom{2n-s}{n}\binom{m+s}{m} = \binom{2n+m+1}{n},
\end{equation}
which is proved by equating coefficients of $x^n$ in the binomial expansions of $(1-x)^{-(n+1)}(1-x)^{-(m+1)}$ and $(1-x)^{-(n+m+2)}$. Thus
\begin{align}
K_{n+1}&=\frac{6\pi^2}{16^{n+1}}\sum_{r=0}^n \binom{2n-r}{n}\binom{2n+r+4}{n}\binom{r+3}{r}\nonumber\\
&=\frac{2\pi^2}{16^{n+1}}\binom{2n}{n}\binom{2n+3}{n}(2n+1)(n+3),
\end{align}
using computer algebra in Maple to evaluate the sum on $r$. Simplifying, 
and replacing $n+1$ by $n$ throughout, we find
\begin{equation}
K_n = \frac{2\pi^2 n^2}{16^n}\binom{2n-1}{n}\binom{2n+1}{n}.
\end{equation}

Second, let $g(x)=\vartheta(x)x^{-\gamma}e^{-1/x}$ and consider 
\begin{equation}
K_\gamma := \int_0^\infty \omega^3\, |\hat{g}(\omega)|^2\,\d \omega ,
\end{equation}
for $\gamma>1$, which is needed to ensure that the transform exists.
Noting that
\begin{equation}
|\hat{g}(\omega)|^2 = \int_0^\infty\int_0^\infty \d x\,\d y\, (xy)^{-\gamma}
e^{-(1/x+1/y)} \cos\omega(x-y),
\end{equation}
the $\omega$ integral can be performed under the $x$ and $y$ integrals to give
\begin{align}
K_\gamma &= \lim_{\epsilon\to 0+} \Re  \int_0^\infty\int_0^\infty \d x\,\d y\, \frac{6(xy)^{-\gamma}
e^{-(1/x+1/y)}}{(x-y-i\epsilon)^4} \non\\
&=\lim_{\epsilon\to 0+} \Re  \int_0^\infty\int_0^\infty \d x\,\d y\, \frac{6(xy)^{-\gamma-4}
	e^{-(1/x+1/y)}}{(y^{-1}-x^{-1}-i\epsilon (xy)^{-1})^4}.
\end{align}
Now change variables so that $x^{-1}=r^2\sin^2\theta$,  $y^{-1}=r^2\cos^2\theta$,
for which the Jacobian factor is $2^5/(r^5\sin^3 2\theta)$. Then
\begin{align}
K_\gamma &=\frac{6}{2^{2\gamma+3}} \left(\int_0^\infty \d r\, r^{4\gamma+3}e^{-r^2}\right)
\lim_{\epsilon\to 0+} \Re \int_0^{\pi/2} \d\theta \, \frac{(\sin^2 2\theta)^{\gamma+5/2}}{(\cos 2\theta - \frac{i}{4}\epsilon\sin^2 2\theta)^4} \\
&= \frac{6\Gamma(2\gamma+2)}{2^{2\gamma+5}}
\lim_{\epsilon\to 0+} \Re \int_{-1}^{1} \d c \, \frac{(1-c^2)^{\gamma+2}}{(c - \frac{i}{4}\epsilon (1-c^2))^4} ,
\end{align}
factorising the integral, evaluating the $r$-integral and changing variables 
to $c=\cos^2 2\theta$ in the $\theta$-integral. The regulating term in the denominator may be replaced simply by $i\epsilon$. Successively integrating
by parts three times,
\begin{equation}
K_\gamma = \frac{\Gamma(2\gamma+2)}{2^{2\gamma+5}}
\lim_{\epsilon\to 0+} \Re \int_{-1}^{1} \d c \, 
\frac{1}{c - i\epsilon}
\frac{\d^3}{\d c^3} (1-c^2)^{\gamma+2}
\end{equation}
because the boundary terms vanish. In general, one has
\begin{equation}
\frac{\d^3}{\d c^3}h(c^2)= 12 c h''(c^2) + 8c^3 h'''(c^2)
\end{equation} 
so the $\epsilon\to 0^+$ limit may be taken, giving
\begin{equation}
K_\gamma = \frac{\Gamma(2\gamma+2)}{2^{2\gamma+4}}
 \int_{0}^{1} \d c \, 4(\gamma+2)(\gamma+1)\left((3+2\gamma)(1-c^2)^\gamma - 2\gamma (1-c^2)^{\gamma-1}\right) .
\end{equation}
The integral may be evaluated in terms of $\Gamma$-functions,
\begin{align}
K_\gamma &= \frac{\Gamma(2\gamma+2)}{2^{2\gamma+2}}  (\gamma+2)(\gamma+1)\left((3+2\gamma)2^{2\gamma}\frac{\Gamma(\gamma+1)^2}{\Gamma(2\gamma+2)}-   2\gamma 2^{2\gamma-2}\frac{\Gamma(\gamma)^2}{\Gamma(2\gamma)}\right) \non\\
&= \frac{\Gamma(\gamma)^2}{2^{2\gamma+2}} 
(\gamma+2)(\gamma+1)\left((3+2\gamma)2^{2\gamma}\gamma^2 -   2\gamma 2^{2\gamma-2}(2\gamma)(2\gamma+1)\right) \non\\
&=\frac{\Gamma(\gamma+1)^2}{4} 
(\gamma+2)(\gamma+1)\left((3+2\gamma) -   (2\gamma+1)\right) \non\\
&=\frac{\Gamma(\gamma+1)\Gamma(\gamma+3)}{2} .
\end{align}


\end{document}